\documentclass[twocolumn]{aastex62}

\usepackage{amsmath}
\usepackage{ulem}

\graphicspath{{./}{fig/}}

\accepted{2021 January 13, ApJ}

\shorttitle{Hi-res emission spectra from cloudy 3D GCMs}
\shortauthors{Harada et al.}

\begin{document}

\title{Signatures of Clouds in Hot Jupiter Atmospheres: \\ Modeled High Resolution Emission Spectra from 3D General Circulation Models}

\correspondingauthor{Caleb K. Harada}
\email{charada@berkeley.edu}

\author[0000-0001-5737-1687]{Caleb K. Harada}\altaffiliation{NSF Graduate Research Fellow}
\affiliation{Department of Astronomy, University of California Berkeley, University Drive, Berkeley, CA 94720, USA}
\affiliation{Department of Astronomy, University of Maryland, 4296 Stadium Drive, College Park, MD 20742, USA}
\affiliation{Department of Astronomy, University of Michigan, 1085 South University Avenue, Ann Arbor, MI 48109, USA}

\author[0000-0002-1337-9051]{Eliza M.-R. Kempton}
\affiliation{Department of Astronomy, University of Maryland, 4296 Stadium Drive, College Park, MD 20742, USA}

\author[0000-0003-3963-9672]{Emily Rauscher}
\affiliation{Department of Astronomy, University of Michigan, 1085 South University Avenue, Ann Arbor, MI 48109, USA}

\author[0000-0001-8206-2165]{Michael Roman}
\affiliation{School of Physics and Astronomy, University of Leicester, University Road, Leicester LE1 7RH, UK}
\affiliation{Department of Astronomy, University of Michigan, 1085 South University Avenue, Ann Arbor, MI 48109, USA}

\author[0000-0003-0217-3880]{Isaac Malsky}
\affil{Department of Astronomy, University of Michigan, 1085 South University Avenue, Ann Arbor, MI 48109, USA}

\author{Marah Brinjikji}
\affiliation{School of Earth and Space Exploration, Arizona State University, PO Box 871404, Tempe, AZ 85287, USA}
\affiliation{Department of Astronomy, University of Michigan, 1085 South University Avenue, Ann Arbor, MI 48109, USA}

\author{Victoria DiTomasso}
\affiliation{Center for Astrophysics $|$ Harvard \& Smithsonian, 60 Garden Street, Cambridge, MA 02138, USA}
\affiliation{Department of Astronomy, University of Michigan, 1085 South University Avenue, Ann Arbor, MI 48109, USA}

\begin{abstract}
    Observations of scattered light and thermal emission from hot Jupiter exoplanets have suggested the presence of inhomogeneous aerosols in their atmospheres. 3D general circulation models (GCMs) that attempt to model the effects of aerosols have been developed to understand the physical processes that underlie their dynamical structures. In this work, we investigate how different approaches to aerosol modeling in GCMs of hot Jupiters affect high-resolution thermal emission spectra throughout the duration of the planet's orbit. Using results from a GCM with temperature-dependent cloud formation, we calculate spectra of a representative hot Jupiter with different assumptions regarding the vertical extent and thickness of clouds. We then compare these spectra to models in which clouds are absent or simply post-processed (i.e., added subsequently to the completed clear model). We show that the temperature-dependent treatment of clouds in the GCM produces high-resolution emission spectra that are markedly different from the clear and post-processed cases---both in the continuum flux levels and line profiles---and that increasing the vertical extent and thickness of clouds leads to bigger changes in these features. We evaluate the net Doppler shifts of the spectra induced by global winds and the planet's rotation and show that they are strongly phase-dependent, especially for models with thicker and more extended clouds. This work further demonstrates the importance of radiative feedback in cloudy atmospheric models of hot Jupiters, as this can have a significant impact on interpreting spectroscopic observations of exoplanet atmospheres. 
\end{abstract}


%
%
\section{Introduction} \label{sec:intro}

Hot Jupiters are perhaps the most extensively studied collection of planets beyond our Solar System. Yet our understanding of the physical processes that govern hot Jupiters is still incomplete. They reside under extreme conditions that are not accessible to Solar System studies, likely giving them a myriad of unfamiliar physical properties. With semi-major axes of a few percent of an AU, they are highly irradiated and are believed to be tidally locked with rotation rates synchronous with their orbits \citep[e.g.,][]{rasio+1996}. Thus, one side of the planet is constantly bathed in intense stellar radiation, while the opposite side is left in perpetual darkness. This is expected to lead to atmospheric structure and dynamics in a regime that has yet to be examined up-close.

Aerosols (i.e., clouds and hazes) are likely an important consideration in studies of hot Jupiter atmospheres because of their ability to scatter and absorb radiation beyond what is expected for a gaseous atmosphere alone \citep{heng+2013}. Aerosols are prominent features of Solar System planets with thick atmospheres \citep[e.g.,][]{west+1986} and have been proposed to explain several unexpected trends in observations of exoplanet atmospheres \citep{marley+2013, sing+2016, crossfield+2017}. As an additional source of opacity, clouds can dampen absorption features in exoplanet transmission spectra \citep[e.g.,][]{deming+2013, kreidberg+2014b, kreidberg+2018, sing+2016, stevenson2016} and weaken the thermal emission from a planet's nightside \citep[e.g.,][]{stevenson+2014, stevenson+2017, keating+2019, demory+2013}. Moreover, asymmetries in reflected light phase curves and unexpectedly high albedos of hot Jupiters have been attributed to the reflective properties of clouds \citep[e.g.,][]{demory+2011, demory+2013, esteves+2015, hu+2015,  parmentier+2016}.

However, the physical characteristics of exoplanetary aerosols remain largely undetermined. Depending on their chemical composition, vertical extent, particle size, and transport through the atmosphere, aerosols can interact with the outgoing thermal emission from the planet's interior and the incoming stellar irradiance to varying degrees. They can therefore alter heating rates and the overall energy balance of an otherwise clear atmosphere, changing the chemical, thermal, and dynamical structure of the initial atmosphere through radiative feedback \citep{heng+2013, lee+2016, lee+2017, lines+2018, roman+2019}. This can have important observable consequences. Since a planet's emission is set by the chemical, thermal, and dynamical structure of the atmosphere, aerosols likely have a considerable influence over the flux we observe from the planet. In order to correctly interpret observations and characterize hot Jupiter atmospheres, it is therefore necessary to assess the impact of clouds (as one source of aerosols) on observable properties.

The complexity of physics and range of scales associated with cloud formation make cloudy hot Jupiter atmospheres challenging to model. Trade-offs must be made between capturing the relevant physical processes involved, exploring a large range of parameter space, and minimizing computational expense. This has led to an assortment of numerical approaches to modeling clouds. Recent efforts have encompassed both 1D \citep[e.g.,][]{morley+2015, gao+2018, powell+2018, ormel+2019} and 3D \citep[e.g.,][]{oreshenko+2016, boutle+2017, roman+2017, roman+2019, lines+2019} frameworks, with the most comprehensive and self-consistent models including spontaneous aerosol formation with radiative feedback, microphysics, and evolution of clouds in their predictions of winds, temperatures, and cloud distributions \citep[e.g.,][]{lee+2016, lee+2017, lines+2018, lines+2019}. Other models have prescribed and radiatively inactive clouds whose distribution is determined by computing a clear (cloud-free) atmosphere and comparing temperature profiles to condensation curves of different cloud species \citep[e.g.,][]{parmentier+2016, kataria+2016}. The latter post-processing method is more computationally efficient, but neglects the feedback effect of aerosols on the thermal structure of the atmosphere. The diversity of cloud modeling approaches has been necessary to determine the minimum level of completeness needed to robustly interpret observations without oversimplifying or ignoring important physics or incorporating extraneous (and computationally expensive) physical processes that have little impact on observables.

Since hot Jupiters are inherently three-dimensional objects that are subject to intense stellar radiation on a single hemisphere, their atmospheric structures and cloud distributions are expected to be non-uniform. However, exoplanets cannot be spatially resolved, and their surface features cannot be directly probed through imaging. Nonetheless, indirect spectroscopic techniques have had great success in characterizing hot Jupiter atmospheres. For transiting planets, the most commonly used techniques involve low- to mid-resolution spectra ($R \lesssim 10,000$) obtained during transit and secondary eclipse, which have constrained chemical abundances and the presence of clouds in exoplanet atmospheres \citep[e.g.,][]{kreidberg+2014a, kreidberg+2018, stevenson+2016, benneke+2019}; and full-orbit phase curves, which have provided constraints on their 3-D temperature structures \citep[e.g.,][]{demory+2013, stevenson+2014, stevenson+2017}.

High-resolution spectroscopy, or HRS ($R \gtrsim 10,000$), has recently emerged as a complementary approach to characterizing exoplanet atmospheres. Not only can HRS be used to access a planet's atmospheric composition, structure, and clouds, but it can uniquely be used to directly detect atmospheric dynamics, such as global wind patterns and planet rotation rates \citep{miller-ricci-kempton+2012,  showman+2013, kempton+2014, rauscher+2014, snellen+2014, louden+2015, brogi+2016, zhang+2017, flowers+2019}, as it is sensitive to the depth, shape, and position of individual spectral lines. HRS is a particularly attractive technique for characterizing hot Jupiters, which orbit close-in to their host stars, because the spectral lines undergo large Doppler-shifts during the planet's orbit. The stellar spectrum and local telluric absorption remain relatively stationary, while the orbital velocity of the planet allows its spectrum to be isolated. While HRS is a powerful tool for studying higher-order atmospheric effects like global winds and rotation, HRS observations are challenging with current technology because they require more photons than low- to mid-resolution spectroscopy to achieve adequate signal-to-noise. However, in the forthcoming era of extremely large ground-based telescopes with HRS capabilities, this issue is likely diminished and high-resolution observations will become more accessible and routine.

Properties of an exoplanet's atmosphere can be empirically constrained by comparing observed spectra to radiative transfer models. With general success, this has typically been done using one-dimensional models that assume the same atmospheric temperature profile and chemical composition along any radial direction \citep[e.g.,][]{morley+2015, gao+2018}. However, the assumption of treating the atmosphere as a 1D global average can lead to biases that ultimately result in incorrect atmospheric constraints \citep{feng+2016, line+2016}. Whereas these models lack key information on thermal structure and dynamics, 3D atmospheric models can treat longitudinal, latitudinal, and vertical variations in temperature, wind, and composition in a more physically-motivated manner. 

A number of 3D general circulation models (GCMs) have been developed to investigate detailed atmospheric structure and dynamics of hot Jupiters, most of which arrive at qualitatively similar results \citep[e.g.,][]{showman+2009, dobbs-dixon+2010, rauscher+2010, heng+2011, mayne+2014}. Though each model contains different assumptions regarding boundary conditions and the complexity of physics involved, all make three key predictions for tidally-locked hot Jupiters: the presence of a super-rotating equatorial jet in the same direction as the planet's rotation; an offset of the planet's hottest region eastward of the sub-stellar longitude; and a strong day-night temperature difference at low pressures along with more homogeneous temperatures at higher pressures. Because these factors can influence the characteristics of individual spectral lines, accurate predictions of high-resolution spectra rely on the self-consistent 3D treatment of temperatures and winds in exoplanet atmospheres \citep[e.g.,][]{miller-ricci-kempton+2012, zhang+2017}; and it has recently been shown that the use of these 3D predictions directly within the data analysis enhances the detection of the planet's signal \citep{flowers+2019}.

In the recent work of \citet{zhang+2017}, a 3D circulation model was post-processed with a radiative transfer solver to incorporate 3D temperature, pressure, and wind information into the calculation of hot Jupiter thermal emission spectra. Using the GCM of \citet{rauscher+2012} and a modified version of the radiative transfer codes implemented in \citet{miller-ricci+2009} and \citet{miller-ricci-kempton+2012}, the authors predicted spectra as a function of orbital phase for HD~209458b, WASP-43b, and HD~189733b. Assuming edge-on and tidally-locked orbits, they  showed that net Doppler shifts were present  in the disk-integrated planet spectra, which varied in a quasi-sinusoidal pattern over the course of each planet's orbit as the substellar hotspot rotated toward and away from the observer. They found that net Doppler shifts of order several km/s resulted from the combined effects of winds, rotation, and thermal structure.

As stated previously, clouds are inherently linked to the 3D temperatures and winds in an exoplanet's atmosphere and their radiative response to their environment affects the ensuing cloud distribution. Therefore, it is expected that clouds have a significant effect on high-resolution spectra exoplanets \citep[e.g.,][]{pino+2018}. Recently, \cite{roman+2019} investigated how radiative feedback can impact cloud distributions, atmospheric temperatures, and observable fluxes from hot Jupiters. In a cloudy GCM of moderate complexity, they parameterized condensate clouds in a temperature-dependent framework similar to the aforementioned post-processing approach, but self-consistently allowed clouds to form and evaporate according to the local temperature and pressure conditions throughout the duration of the GCM, rather than only after the final step.  This approach allowed the authors to account for the radiative feedback of the clouds on the local temperatures and atmospheric dynamics, as they evolved during the GCM simulation.

In this study, we aim to assess the thermal and dynamical influence of radiatively active clouds on hot Jupiter atmospheres, as detectable via observations of high-resolution thermal emission spectra. Here, we post-process the 3D cloudy GCM of \cite{roman+2019} with a radiative transfer solver adapted from \cite{zhang+2017} to compute high-resolution spectra that include attenuation effects from inhomogeneous clouds. We test multiple cases with different assumed cloud characteristics and perform cross-correlations to rest-frame spectra (calculated in the same 3D framework) to evaluate the net Doppler shift due to global wind patterns and planet rotation, as a function of orbital phase. This paper is organized as follows. In Section \ref{sec:methods} we summarize the GCM of \cite{roman+2019} and our radiative transfer code. In Section \ref{sec:results} we describe the results of our GCM for various cloud assumptions and resulting thermal emission spectra. Finally, we summarize our main conclusions in Section \ref{sec:conclusion}.

%
%
\section{Methods} \label{sec:methods}

We use 3D models with radiatively active clouds to simulate several possible states that a hot Jupiter atmosphere could have; then we use a detailed radiative transfer routine to post-process the predicted temperature, pressure, velocity, and cloud structures in order to create simulated high-resolution spectra; and lastly we cross-correlate those spectra with their un-Doppler-shifted versions to calculate the net atmospheric Doppler effects. These components are described in the following sections.

\subsection{3D General Circulation Model} \label{sec:methods:gcm}

We compute synthetic emission spectra directly from the GCM simulations previously presented in \citet{roman+2019} and summarized as follows. \citet{roman+2019}, henceforth RR19, used a GCM \citep{rauscher+2010,rauscher+2012,RauscherMenou2013} that solves the primitive equations of meteorology, coupled to a double-gray, two-stream radiative transfer scheme based on \citet{toon+1989}. Atmospheric temperatures, winds, radiative fluxes, and idealized cloud distributions were simulated for a hot Jupiter with a mass, radius, and irradiance temperature based on Kepler-7b \citep[][model input parameters listed in Table~\ref{tab:planet_params}]{latham+2010, demory+2011, demory+2013}.

\begin{deluxetable*}{lCCl}[t]
\tablecaption{GCM Planet Parameters \label{tab:planet_params}}
\tablehead{
\colhead{Parameter} &
\colhead{\hspace{1.2cm}Value}\hspace{1.2cm} &
\colhead{\hspace{1.2cm}Units}\hspace{1.2cm} &
\colhead{\hspace{0.5cm}Reference}\hspace{0.5cm}
}
\startdata
Planet radius, $R_\text{p}$ & 1.128 \times 10^8 & $\text{m}$ & \citet{demory+2011} \\
Gravitational acceleration, $g$ & 4.17 & \text{m}\,\text{s}^{-2} & \citet{demory+2011} \\
Planet rotation rate\tablenotemark{a}, $\Omega$ & 1.49\times10^{-5} & \text{s}^{-1} & \citet{roman+2019} \\
Incident flux at substellar point, $F_{\downarrow\text{vis,irr}}$ & 1.589\times10^6 & \text{W}\,\text{m}^{-2} & \citet{demory+2011} \\
Internal heat flux, $F_{\uparrow\text{IR,int}}$ & 2325 & \text{W}\,\text{m}^{-2} & \citet{roman+2019} \\
Semi-major axis, $a$ & 0.061 & \text{AU} & \citet{esteves+2015} \\
Orbital period, $P$ & 4.89 & \text{day} & \citet{esteves+2015} \\
Equilibrium temperature, $T_\text{eq}$ & 1630 & \text{K} & \citet{esteves+2015} \\
\enddata
\tablenotetext{a}{The rotation rate is assumed to be synchronous with the planet's orbit.}
\tablecomments{This planet is modeled after Kepler-7b; more GCM parameters can be found in \citet{roman+2019}.}
\end{deluxetable*}

The GCM's radiative transfer calculations include the effects of aerosol scattering and absorption with condensate clouds parameterized as Mie scatterers with horizontal spatial distributions solely dependent on the temperature field. In this idealized approach, aerosol scattering was included where the atmospheric temperatures fell below the expected condensation temperature for given condensable species. Four species were included: MnS, Al$_2$O$_3$, Fe, and MgSiO$_3$.  The scattering properties (i.e., single scattering albedo and asymmetry parameter) were pre-computed for specific cloud species assuming particles were 0.2 $\mu$m in radius at two wavelengths\textemdash 0.65 $\mu$m and 5.0 $\mu$m\textemdash representative of the visible and thermal channels of the double-gray model.  Where clouds of different species overlapped, the scattering properties were averaged, weighted by the relative optical thickness of each species. Aerosol abundances were taken to be proportional to the compositional abundance of the condensing gas with constant volume mixing ratio, but scaled by an adjustable parameter to control the overall optical thickness of the cloud. 

The GCM vertically resolved the atmosphere using a sigma coordinate system, in which the computed temperatures, winds, and clouds were defined at the discrete pressure levels corresponding to these discrete sigma levels, which were ultimately converted into altitude assuming hydrostatic equilibrium. There was no spatial averaging beyond this resolution.

RR19 focused on four cases that explored basic assumptions regarding the cloud thickness and extent.  Half the simulations assumed optically thicker clouds equivalent to condensing 1/10 of the gaseous species; the other simulations assumed thinner clouds with only 1/100 of the gas condensed. For each assumed thickness, two different assumptions regarding the vertical extent of the cloud were explored.  In one pair of simulations, clouds were truncated $\sim$1.4 scale heights (five vertical layers) above the cloud base pressure, representing an atmosphere with weak vertical mixing. These two cases were referred to as the \emph{compact thick} and \emph{compact thin} cases. For contrast, in the other pair of simulations, clouds were allowed to extend to the 0.1 mbar pressure level (temperature permitting and regardless of the base pressure). These were referred to as the \emph{extended thick} and \emph{extended thin} models, and represented atmospheres with more efficient vertical mixing.  

For each of these cloud models, RR19 sought to evaluate the importance of cloud radiative effects in shaping the atmospheric temperatures and fluxes.  As such, they performed two versions of each simulation. In the first version, clouds were included throughout the duration of the 2000-orbit simulation, continually forcing and responding to the temperature field. They referred to this as \emph{active} cloud modeling since the clouds actively influenced the thermal and dynamical structure of the atmosphere.  For each of these active cloud simulations, RR19 ran a contrasting model in which clouds were absent throughout the simulation (i.e., a clear model), and only \emph{post-processed} by comparing condensation curves to the final temperature field. The latter post-processing approach has had precedence in the literature in predicting and interpreting hot Jupiter phase curve data \citep[e.g.,][]{parmentier+2016, kataria+2016}. In RR19 the difference between the active and post-processed cloud cases demonstrated the role of aerosol radiative forcing and feedback in the model.

\subsection{Emission Spectra Radiative Transfer} \label{sec:methods:rt}

We generated thermal emission spectra at high resolution ($R \sim 10^6$) over the 2.308 - 2.314 $\mu$m wavelength range, corresponding to a strong CO band head that has been the target of exoplanet atmospheric characterization with the CRIRES instrument on the VLT \citep[e.g.,][]{de-kok+2013, de-kok+2014, schwarz+2015, brogi+2013, brogi+2014, brogi+2016}. From each GCM output from RR19, we calculated spectra in the planet's rest frame, with and without including Doppler effects from atmospheric motion (winds and rotation), at 24 evenly-spaced orbital phases, corresponding to different viewing geometries. 3D temperatures, winds, and cloud optical depths (computed at 5.0 $\mu$m) throughout the planet's atmosphere were taken from the last time step of the GCM simulations.

\begin{figure*}[t]
    \centering
    \includegraphics[width=\textwidth]{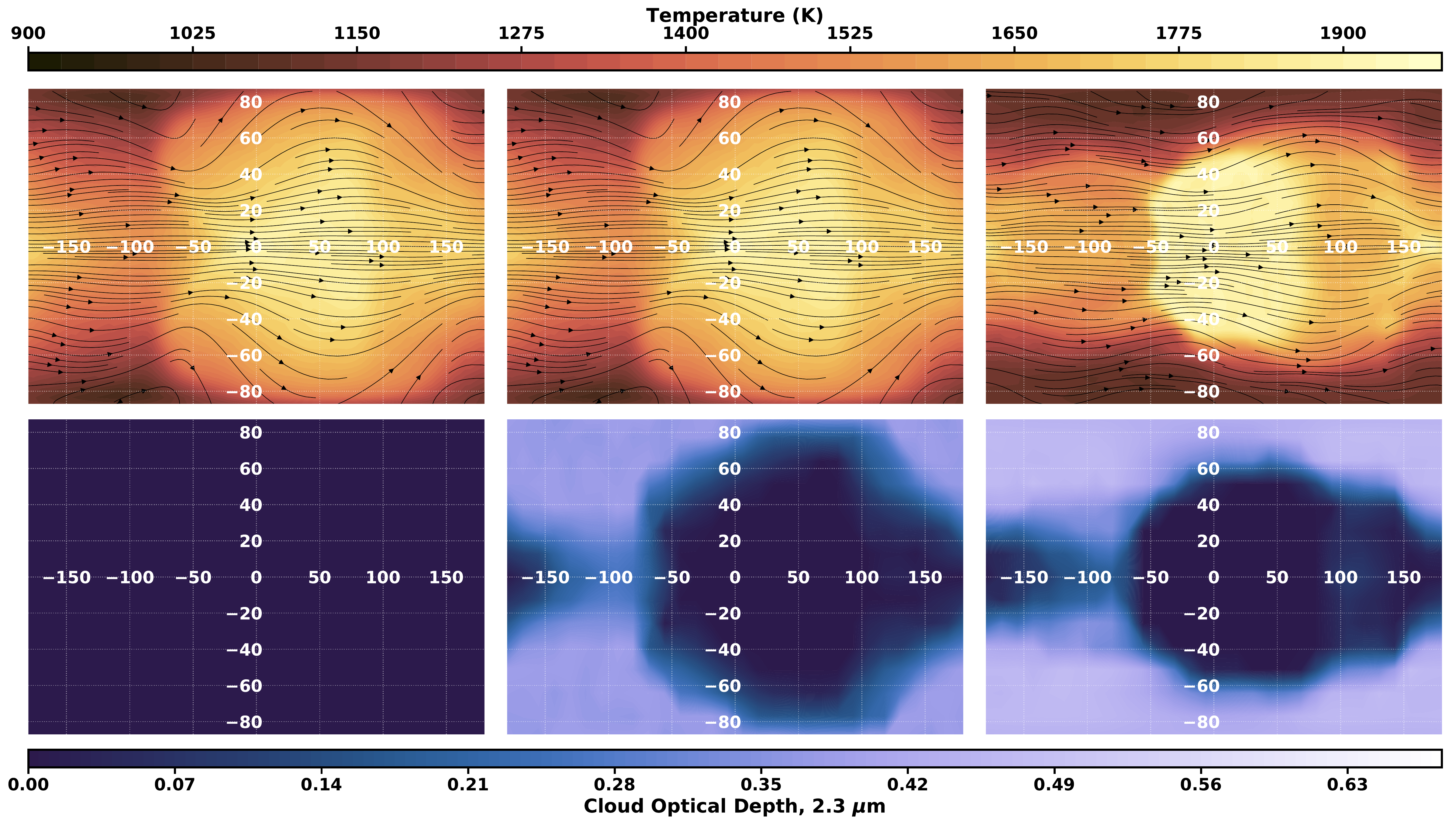}
    \begin{minipage}[]{0.32\textwidth}
        \centering
        (a) Clear
    \end{minipage}
    \begin{minipage}[]{0.32\textwidth}
        \centering
        (b) Post-processed Clouds
    \end{minipage}
    \begin{minipage}[]{0.32\textwidth}
        \centering
        (c) Active Clouds
    \end{minipage}
    \caption{\textit{Top}: Longitude-latitude maps (centered on the substellar point) of simulated temperatures and winds from the clear GCM (a), post-processed cloud GCM (b), and active cloud GCM (c) from \citet{roman+2019}. \textit{Bottom}: Corresponding longitude-latitude maps of infrared (2.3 $\mu$m) cloud optical depth. Both cloudy models assume extended thick cloud properties. The temperatures and wind vectors correspond approximately to the infrared photosphere pressure level of the clear model ($\sim$26 mbar) and the IR cloud optical depths are integrated vertically above the photosphere. The post-processed cloud GCM has an identical thermal and dynamical structure to the clear GCM, while the active cloud GCM has a markedly different underlying structure.}
    \label{fig:temp_maps}
\end{figure*}

In our radiative transfer calculations, we modeled clouds as a source of pure thermal absorption/emission. We note, however, that aerosols can potentially scatter light back into the outgoing beam and partly mitigate the overall observed extinction, particularly if single-scattering albedos are high and scattering phase functions are strongly forward-scattering\footnote{Though our radiative transfer post-processing code here does not include scattering, the double-gray radiative transfer treatment in the GCM \emph{does}.}. Here, because the total cloud distributions are optically thin in the IR (see Figure \ref{fig:temp_maps}), and because MgSiO$_3$, Al$_2$O$_3$, and MnS are near-isotropic scatterers in the IR \citep{roman+2019}, we assume that the effects of non-isotropic scattering are relatively small. Hence we neglect the contribution of scattering to the observed planet flux and leave a more complete scattering treatment to future work\footnote{Malsky et al. (in prep) will investigate the effects of non-isotropic scattering on hot Jupiter emission spectra using a two-stream radiative transport approximation.}.

Following \citet{zhang+2017}, we calculated the emission spectrum from the GCM output by summing the intensity of all emergent line-of-sight rays from the planet's visible side. The emergent intensity of each ray was calculated using the standard radiative transfer equation in the thermal emission approximation:
\begin{equation}\label{eq:rt}
    I_\lambda = B_\lambda(\tau_\text{max}) e^{-\tau_\text{max}} + \int_{\tau_\text{max}}^0 B_\lambda(\tau) e^{-(\tau_\text{max} - \tau)} \,d\tau
\end{equation}
where $\tau$ is the slant optical depth along the sight line (implicitly wavelength-dependent; defined below), $B_\lambda(\tau)$ is the (Planckian) source function evaluated at the local temperature, and $B_\lambda(\tau_\text{max})$ is the source function evaluated at the base layer of the atmosphere, defined here as $\sim$100 bar in pressure (RR19). Here, the first term accounts for radiation emitted from the bottom of the atmosphere and attenuated by gas and clouds along the line-of-sight. The second term accounts for radiation emitted and absorbed by each parcel of gas and clouds in the path of the initial light ray.

The slant optical depth is defined as
\begin{equation}\label{eq:tau}
    \tau_\lambda = \int (\kappa_\text{gas} + \kappa_\text{cloud})\, dl
\end{equation}
where $\kappa_\text{gas}$ is the wavelength-dependent gas opacity evaluated at local atmospheric temperature and pressure, $\kappa_\text{cloud}$ is the local cloud opacity, and $dl$ is the line-of-sight path length through each grid cell encountered by the ray. We assume solar composition gas in local thermochemical equilibrium and that cloud opacities are gray (i.e., wavelength independent) over the narrow wavelength range of the calculations pursued in this work.

For these LOS calculations, we used altitude as the vertical coordinate after interpolating the GCM output to a grid of constant $\Delta$-altitude. We note that because different parts of the planet had different scale heights, not all of the T-P profiles extended upward to the same altitude---the local opacities were set to zero at altitudes where the pressure dropped below the upper boundary of the atmosphere in the GCM simulations ($\sim 57~\mu$bar; RR19).

For Equation \ref{eq:tau}, we converted the cloud optical depth $d\tau_\text{cloud}$ of each condensate species in a grid cell of the GCM (expressed at 5.0 $\mu$m by RR19) into an effective extinction opacity at 2.3 $\mu$m with the following transformation:
\begin{equation} \label{eqn_clouds}
    \kappa_\text{cloud} = \frac{d\tau_\text{cloud}}{ds} \frac{Q_{2.3\mu\text{m}}}{Q_{5.0\mu\text{m}}}
\end{equation}
where the ratio of the extinction efficiencies (${Q_{2.3\mu\text{m}}} / {Q_{5.0\mu\text{m}}}$) is used to convert extinction at 5.0 $\mu$m to the extinction at 2.3 $\mu$m, and $ds$ is the differential path length defined in the radial direction. The calculation described in Equation~\ref{eqn_clouds} was performed individually for each of the four cloud species from the GCM (given in Table~\ref{tab:clouds}), and then the total cloud opacity $\kappa_\text{cloud}$ was taken as the sum of the single-species opacities.

\begin{deluxetable*}{lCCCC}[t]
\tablecaption{Cloud Scattering Parameters \label{tab:clouds}}
\tablehead{
\colhead{Parameter} &
\colhead{\hspace{0.5cm}$\text{MgSiO}_3$}\hspace{0.5cm} &
\colhead{\hspace{0.5cm}$\text{Fe}$}\hspace{0.5cm} &
\colhead{\hspace{0.5cm}$\text{Al}_2\text{O}_3$}\hspace{0.5cm} &
\colhead{\hspace{0.5cm}$\text{MnS}$}\hspace{0.5cm}
}
\startdata
Molecular weight, $\mu_g$ [g\,mol$^{-1}$] & 100.4 & 55.8 & 102.0 & 87.0 \\
Mole fraction, $\chi_g$ & 3.26\times10^{-5} & 2.94\times10^{-5} & 2.77\times10^{-6} & 3.11\times10^{-7} \\
Particle density, $\rho$ [g\,cm$^{-3}$] & 3.2 & 7.9 & 4.0 & 4.0 \\
Refractive index, $\tilde{n}$ & 1.5 + i(4 \times 10^{-4}) & 4.1 + i8.3 & 1.6 + i(2 \times 10^{-2}) & 2.6 + i(1 \times 10^{-9}) \\
\hline
Extinction efficiency ($\lambda=2.3\,\mu$m), $Q_{2.3\mu\text{m}}$ & 0.07 & 1.25 & 0.12 & 0.56 \\
Extinction efficiency ($\lambda=5\,\mu$m), $Q_{5.0\mu\text{m}}$ & 0.01 & 0.16 & 0.02 & 0.02 \\
Optical depth per bar (thin clouds), $\tau/\Delta P$ & 29.0 & 105.1 & 3.4 & 1.5 \\
Optical depth per bar (thick clouds), $\tau/\Delta P$ & 290.0 & 1051.5 & 34.3 & 15.3 \\
\enddata
\tablecomments{All scattering parameters are shown for 2.3 $\mu$m, unless otherwise specified. Optical depths per bar ($\tau / \Delta P$) were calculated from Equations (1) and (2) in \citet{roman+2019}, based on the computed extinction efficiencies ($Q_e^{(\lambda)}$) and assumed mole fractions ($\chi_g$), molecular weights ($\mu_g$), particle radii ($r$), particle densities ($\rho$), planet surface gravity ($g$), and vapor condensation fractions ($f=0.01$ for thin clouds and $f=0.1$ for thick clouds) in an atmosphere of Jovian mean molecular weight. The complex refractive indices ($\tilde{n}$) used to calculate $Q_e^{(\lambda)}$ were taken from \citet{kitzmann+2018}, and $\chi_g$ and $\rho$ were taken from \citet{roman+2019}, with relevant references therein.}
\end{deluxetable*}

Values for the extinction efficiencies were calculated using spherical particles and the indices of refraction cited in Table \ref{tab:clouds}. Following \citet{roman+2019}, we computed these parameters using the Mie scattering code developed by M.I. Mishchenko \citep{de-rooij+1984, mishchenko+1999}, assuming a log-normal particle size distribution with an effective mean radius of 0.2 $\mu$m and variance of 0.1 $\mu$m. Further details on the assumed aerosol scattering properties in the GCM can be found in RR19.

We assumed that the planet's orbital radial velocity can be readily isolated from observational data, and we therefore ignored its contribution to the total motion along the line-of-sight in our calculations. Following \citet{zhang+2017}, we accounted for the remaining local line-of-sight velocity as\footnote{The first term of this equation differs from Equation (3) in \citet{zhang+2017}, due to an error uncovered in that expression. We tested the effect of this correction and found that it has negligible impact on our results.}
\begin{equation}\label{eq:v_los}
\begin{split}
        v_\text{LOS} = & -u\sin(\phi + \varphi) \\
        & - v\cos(\phi + \varphi)\sin(\theta) \\
        & + w\cos(\phi + \varphi)\cos(\theta) \\
        & - \Omega(R_\text{p} + z)\sin(\phi + \varphi)\cos(\theta).
\end{split}
\end{equation}
where the first term represents the contribution from the zonal (i.e., east-west) wind speed $u$, the second term accounts for the meridional (i.e., north-south) wind speed $v$, the third term accounts for the vertical wind speed $w$, and the fourth term accounts for the planet's rotation. In the planet's reference frame $\phi$ is longitude, $\theta$ is latitude (both fixed to the planet's rotation), and $\varphi$ is the phase of the orbit. In the final term, $\Omega$ is the angular rotation speed of the planet (in radians per second; assumed to have been tidally locked into synchronous rotation), $R_\text{p}$ is the radius of the planet at the base of the atmosphere, and $z$ is the vertical height in the atmosphere above $R_\text{p}$. Assuming an edge-on orbit, we defined the substellar point as $\phi=\theta=0$, with transit geometry corresponding to an orbital phase of $\varphi=0$.

Our radiative transfer accurately accounted for the 3D geometry of the atmosphere, following the methods outlined initially in \citet{miller-ricci-kempton+2012} and \citet{zhang+2017}. We self-consistently incorporated the effects of line-of-sight motion by evaluating the local opacities (in Equation \ref{eq:tau}) at their Doppler shifted wavelengths according to
\begin{equation}\label{eq:doppler}
    \lambda = \lambda_0 \Big( 1 - \frac{v_\text{LOS}}{c} \Big)
\end{equation}
where $\lambda_0$ is the rest-frame (i.e., unshifted) wavelength and $c$ is the speed of light.

We calculated a spectrum for a particular orbital phase by dividing the visible hemisphere of the planet into 2304 individual cells, determined by the latitude-longitude grid of the GCM output, and propagating a light ray along the line-of-sight through the atmosphere at each cell according to Equation \ref{eq:rt}. The disk-integrated flux from the planet was the sum of all emergent light ray intensities weighted by the solid angle subtended by each grid cell.

%
%
\section{Results \& Discussion} \label{sec:results}

In this section, we first review the main outcomes of the cloud-free, post-processed cloud, and the active cloud GCMs \citep[presented previously in][]{roman+2019}, then discuss in detail the resulting emission spectra we calculated from each GCM output. Our remaining discussion will focus on the \emph{extended thick} cloud models, since thicker and more extended clouds tended to most drastically affect the thermal structure of the atmosphere, and hence the resulting thermal emission spectra. Though the atmospheric structure and subsequent spectra from our other cloudy models also exhibited differences from the clear GCM results, these dissimilarities were most apparent in the extended thick cloud model.  Additional figures showing our results for all four cloud implementations (extended thick, extended thin, compact thick, and compact thin) are included in Appendix \ref{sec:app:figures}.

\begin{figure*}[t]
    \centering
    \includegraphics[width=\textwidth]{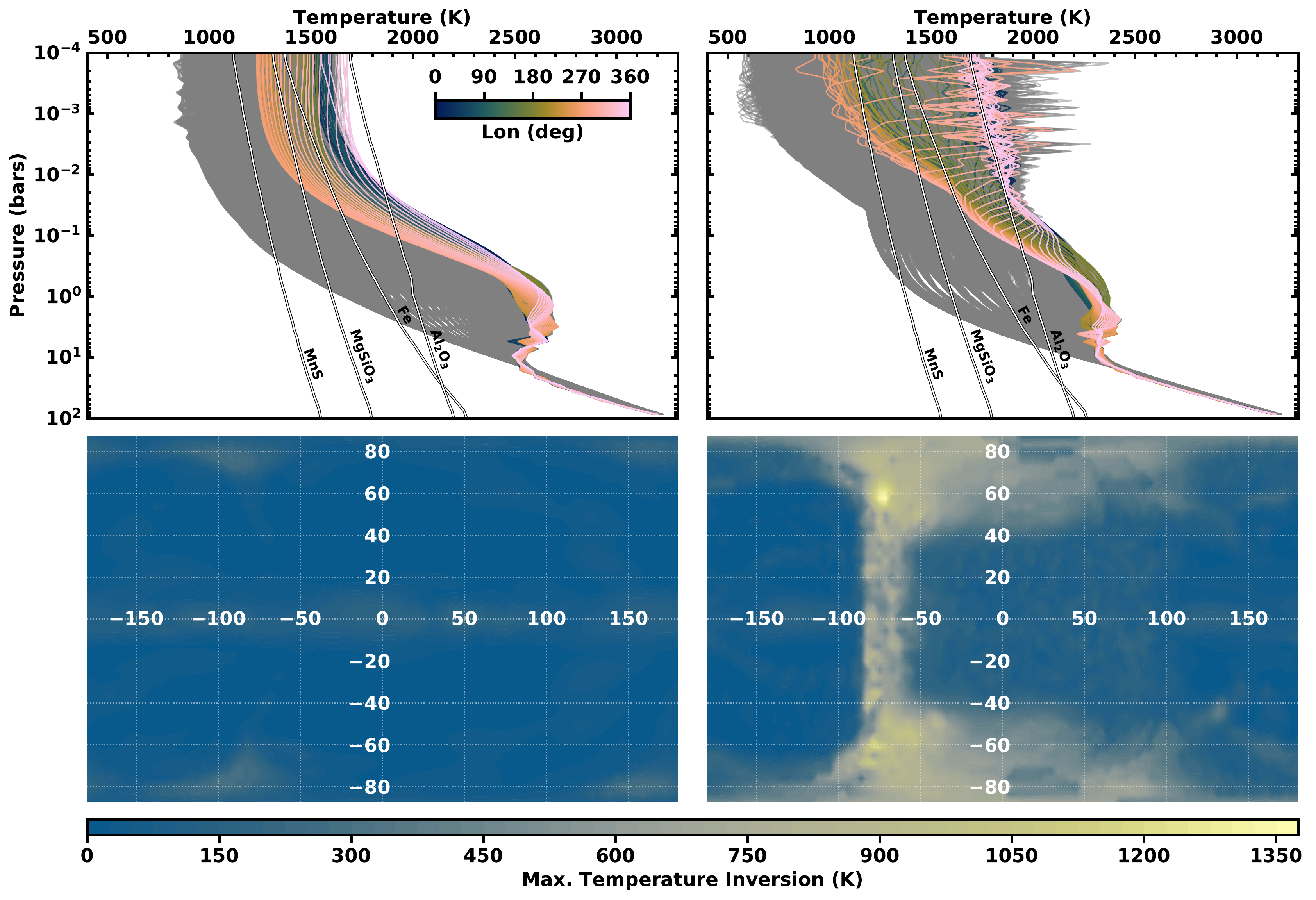}
    \begin{minipage}[]{0.55\textwidth}
        \centering
        (a) Clear/Post-processed Clouds
    \end{minipage}
    \begin{minipage}[]{0.43\textwidth}
        \centering
        (b) Active Clouds
    \end{minipage}
    \caption{\textit{Top}: Vertical temperature-pressure profiles from the clear GCM (a) and active cloud GCM with extended thick clouds (b). The profiles shown in color represent equatorial longitudes and the gray profiles represent all other latitudes and longitudes in the simulated atmosphere. Cloud condensation curves from \citet{roman+2019} are labeled for each species---a cloud base forms at each location where the temperature crosses below a condensation curve. Note that the post-processed cloud GCM by definition has an identical temperature structure to the clear model. Jaggedness in the active cloud temperature profiles (b), seen most notably in the hottest vertical profiles, is the result of numerical noise \citep[see][]{roman+2019} and does not significantly influence our calculated spectra. \textit{Bottom}: Longitude-latitude maps showing the maximum vertical temperature inversion in the clear GCM (a) and active cloud GCM with extended thick clouds (b). These temperature inversions are not a result of numerical noise---they are caused by clouds trapping heat along the terminator and can be seen most clearly in the pressure-temperature profiles for equatorial longitudes around 270 degrees (the western terminator).}
    \label{fig:tp+inversion}
\end{figure*}

\subsection{GCM Outputs} \label{sec:results:gcm}

Longitude-latitude maps of temperatures, winds, and cloud coverage are shown in Figure \ref{fig:temp_maps} for the clear atmosphere and post-processed and active cloud models with extended thick cloud properties. The corresponding vertical thermal structures are shown in Figure \ref{fig:tp+inversion}. Temperature and cloud maps and vertical temperature profiles for the other cloud models are also shown in Figures \ref{fig:all_temp_maps} and \ref{fig:all_tp+inversion} in Appendix \ref{sec:app:figures}.

The clear model reproduced the standard pattern we have come to expect for the hot Jupiter atmospheric circulation regime. The wind structure within the observable atmospheric layers is characterized by eastward flow, dominated by a strong equatorial jet. This jet advects the hottest gas away from the substellar point before it has a chance to efficiently cool, resulting in an eastward offset of the brightest emitting region (see Figure \ref{fig:temp_maps}). 

From the global temperature profiles of the clear model (Figure \ref{fig:tp+inversion}), we can see more typical hot Jupiter behavior. The largest (longitudinal) temperature differences are at the lowest pressure levels (the highest regions of the atmosphere), while temperatures are more homogenized at depth. Large day-night temperature differences exist high in the atmosphere, but this transitions to the main temperature difference being between the equator and poles deeper in the atmospheres. In the clear atmosphere the incoming optical photons are absorbed at deeper atmospheric levels than where the thermal infrared radiation from the planet is emitted to space, meaning that temperatures decrease with pressure. (Deviations from this at $\sim$10 bar are the result of advection dominating the temperature structure at those depths.)

When clouds were included in the GCM, the thickest clouds form near the poles and western terminator, where cooler temperatures allow for MgSiO$_3$, MnS, and Fe to condense. On the dayside, this leads to enhanced cooling at depth as clouds scatter and reflect incoming visible radiation. In contrast, Al$_2$O$_3$, with its relatively higher condensation temperature, also forms clouds over the much of the hot day side and warms the visible atmosphere through absorption of stellar and thermal radiation. The relative importance of these effects depends on the thickness and vertical extent of the clouds, as discussed in RR19. Clouds forming higher in the atmosphere are more effective at altering the albedo and emission, and so the extended clouds have a more significant effect on the atmospheric temperatures.  In general, the increased planetary albedo on the dayside and the increased extinction on the nightside lead to a cooler planet. In all cases, thermal emission is reduced on the nightside relative to a clear or post-processed cloud model, although dayside emission is increased in the thickest cloud models as heat escapes through relatively clearer skies on the warmer dayside. 

An important characteristic of the active cloud models not previously noted in RR19 is the prevalence of thermal inversions along the western terminator (longitude = $-90^\circ$) and dayside polar regions of the planet, as seen in Figure \ref{fig:tp+inversion} (see also Figure \ref{fig:all_tp+inversion} in Appendix \ref{sec:app:figures}). We defined the thermal inversion at a particular location in the atmosphere as the maximum continuous increase in temperature from the bottom to the top of the atmosphere. Stronger thermal inversions tended to form as we transition from models with thinner and more compact clouds to models with thicker and more extended clouds. 

Thermal inversions along the western terminator originate from excess cloud opacity as cool gas is advected eastward from the planet's nightside to the dayside. The cooler nightside temperatures permit cloud formation in the upper atmosphere along the terminator; as these cooler cloudy regions advect past the terminator to the dayside, stellar radiation is scattered and absorbed at the top of the cloud deck. This simultaneously increases heating in these cloudy layers of the atmosphere and cools the lower layers because radiation is impeded from penetrating below the clouds, leading to a thermal inversion. As gas on the planet's western limb continues to move eastward beyond the terminator and absorbs more radiation, the local temperature eventually increases beyond the condensation limit of the clouds. This leads to evaporation of the clouds and an overall decrease in the optical thickness of the atmosphere. As a result, stellar radiation is allowed to penetrate deeper into the atmosphere, returning the temperature profile to a normal, non-inverted one.

Thermal inversions at the planet's higher dayside latitudes can be explained by a similar process. Advection of cooler gas from the nightside of the planet allows clouds to form in the upper atmosphere along the western terminator, which scatter and absorb radiation and prevent heating of the lower layers of atmosphere. However, because stellar radiation is less intense at higher latitudes than near the equator, heating rates in the upper atmosphere fail to rise sufficiently to evaporate clouds and reverse the temperature inversions as the cooler region advects over the dayside.

We note that the presence and strength of thermal inversions ultimately depend on the composition of the atmospheric gas and aerosols. In their modeling, RR19 include Al$_2$O$_3$ and Fe, which tend to strongly absorb solar radiation and lead to thermal inversions \citep{roman+2020}. However, the microphysical models of \citet{gao+2020} suggest strongly scattering silicate clouds should dominate over the absorbing Fe clouds based on arguments of nucleation rates, thus potentially reducing the likelihood of inversions due to clouds. Although a full discussion of the impact of each individual cloud species and assumptions regarding composition are beyond the scope of the this work, the results discussed here nonetheless reveal how clouds with assumed properties can alter the observed spectra.

While these temperature inversions were not discussed explicitly in RR19, we recognize their importance here because of their potential to impact thermal emission spectra via their effect on line shapes (i.e., emission vs. absorption). In our radiative transfer calculations, some regions of the planet can be sources of \emph{emission} lines and other regions can be sources the same lines in \emph{absorption}, both in view at the same time. The implications of this are discussed in the following section.

\subsection{High Resolution Thermal Emission Spectra} \label{sec:results:spectra}

\begin{figure*}[t]
    \centering
    \includegraphics[width=\textwidth]{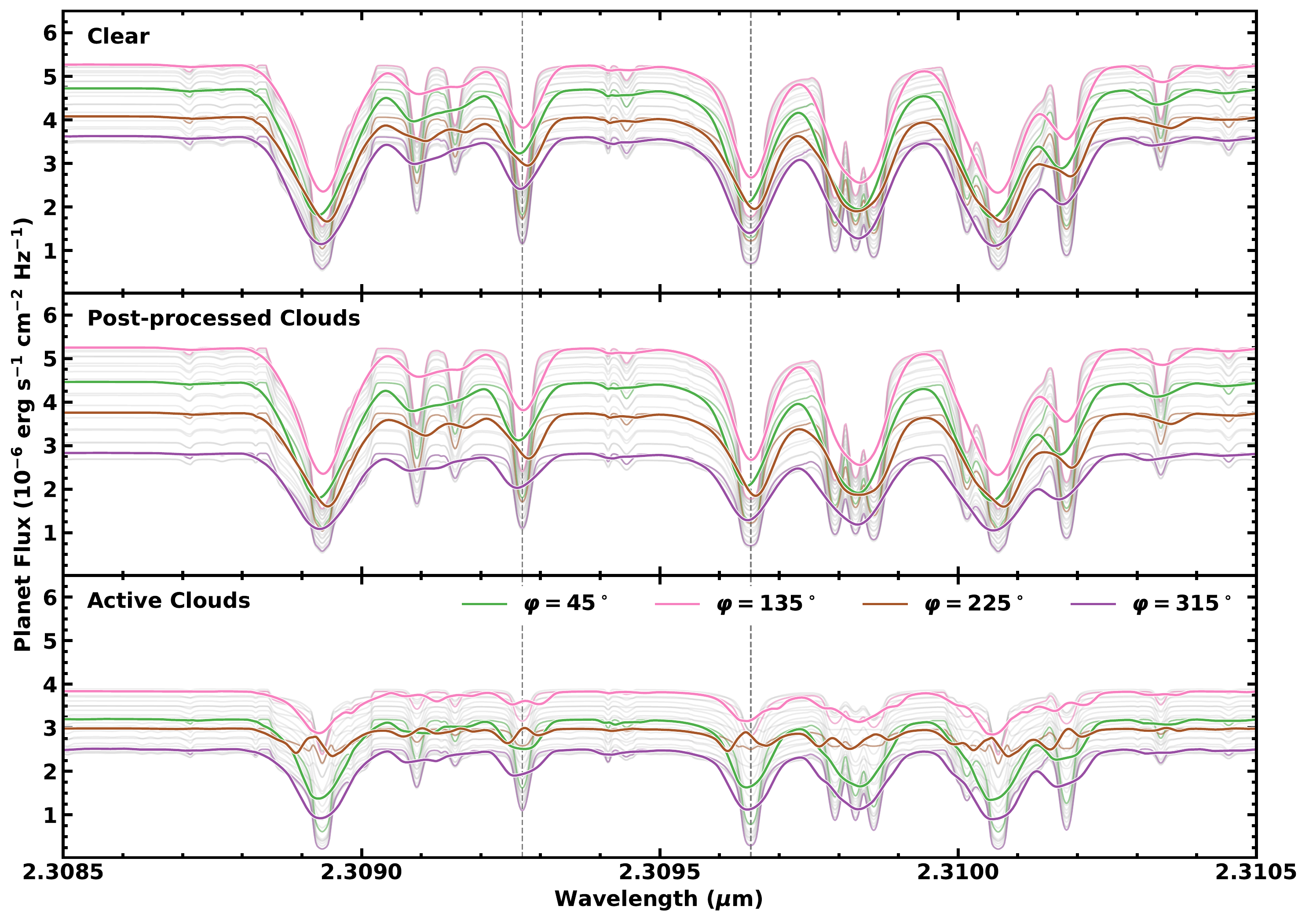}
    \caption{Computed emission spectra from the clear GCM (top), post-processed cloud GCM (center), and active cloud GCM (bottom). Both sets of cloudy spectra are calculated from the models with extended thick clouds. The bold colored spectra include the Doppler shifts from atmospheric motion (winds and rotation), while the lighter colored spectra lack the Doppler effects of these motions---these are shown at orbital phases corresponding to viewing the brightest side of the clear-atmosphere planet ($\varphi=135^\circ$; pink) and intermediate phases at $90^\circ$ intervals. Rest-frame spectra for all other simulated orbital phases are shown in light gray. The vertical dashed lines indicate the rest-frame line centers of prominent H$_2$O features. While the spectra from the clear and post-processed cloud GCMs have similar line shapes, the active cloud GCM spectral lines appear significantly muted, especially on the dayside ($\varphi\sim135^\circ$ and $\varphi\sim225^\circ$), and show weaker day-to-night continuum variations. Note that the brightest and dimmest sides of the planet are visible at orbital phases preceding $\varphi=180^\circ$ and $\varphi=0^\circ$, respectively, as a result of the eastward advection of the planet's hottest region. Note also the presence of small emission-like water features from the brighter side of the active cloud model (e.g., $\sim$2.30928 $\mu$m and $\sim$2.30965 $\mu$m) that correspond to absorption lines in the clear and post-processed cloud models. See Figures \ref{fig:all_spectra1} and \ref{fig:all_spectra2} for additional Doppler-shifted spectra calculated for all 24 evenly-spaced orbital phases.}
    \label{fig:spectra}
\end{figure*}

From the GCM outputs, we calculated two emission spectra at each orbital phase: one with Doppler effects from line-of-sight wind and rotation velocities turned on, and one with Doppler effects turned off. Each rest-frame spectrum was computed as a template for assessing the net Doppler shift produced in the corresponding Doppler-shifted spectrum and to visually examine the effects of Doppler shifts on our models. Figure \ref{fig:spectra} shows our computed rest-frame planet emission spectra and a subset of our Doppler-shifted spectra from the clear, post-processed cloud, and active cloud GCMs. Additional Doppler-shifted spectra for all orbital phases and GCM runs are shown with arbitrary continuum flux in Figures \ref{fig:all_spectra1} and \ref{fig:all_spectra2} in Appendix \ref{sec:app:figures}.

Emission spectra from the post-processed cloudy GCM are qualitatively similar in shape and depth to the spectra from the clear GCM. Because the thermal and dynamical structure of the post-processed atmosphere was identical to that of the clear atmosphere, differences in spectral features are only from gray opacity differences where the cloud deck was optically thick. This is most noticeable in the continuum flux. When the thickest clouds were covering the visible side of the planet, there was a greater deficit in the continuum flux. Since cooler nightside temperatures permitted more cloud formation, the nightside spectra from our post-processed model are the farthest offset from the nightside spectra from the clear model, while the dayside spectra have similar continuum levels in both models. This matches the results presented in \citet{roman+2019} for the broadband thermal phase curves.

The spectra were more drastically affected when we included active clouds in the GCM. Given similar cloud thicknesses above the photosphere in the active and post-processed cases, the additional differences seen in the active cloud case can be primarily attributed to the significant effects clouds have on the temperature field only captured in active modeling. The emission is more strongly altered due to changes to the photosphere temperature and high-altitude thermal inversions as opposed to direct attenuation alone.

First, we note that the continuum fluxes from the active cloud atmosphere are generally lower than the cloud-free and post-processed cloudy atmospheres, which is again consistent with the broadband thermal emission phase curves from \citet{roman+2019}. The variations in continuum flux from the planet's brightest side to dimmest side are also smallest for the active cloud model. This is the result of more widespread and optically thick clouds forming in the active cloud model (see Figure \ref{fig:temp_maps}), which act as an additional gray opacity source that attenuates the continuum flux. The enhanced cloud cover also increases the planetary albedo, which reduces the planet's equilibrium temperature.

More interestingly, the combined effects from additional gray cloud opacities and different thermal structure and dynamics give the spectra from these models features that are distinct from the clear and post-processed models in terms of continuum flux, absorption line depth, and Doppler effects (for the latter, see Section \ref{sec:results:doppler}). A particularly notable result is the flattening of absorption features in the dayside emission spectra from the active cloud model. Recall that the active cloud atmosphere has strong, inhomogeneous temperature inversions (Section \ref{sec:results:gcm}). Because emission features arise from regions of the visible side of the planet that exhibit temperature inversions along the line-of-sight, these emission features overlap with and subsequently cancel out absorption features (of the same line) from regions with non-inverted T-P profiles. This effectively flattens out the features we see in the emission spectra at certain orbital phases.

Moreover, because the absorption/emission lines originate from different regions of the planet, they are created by light rays that traveled along different paths through the atmosphere, and hence encountered grid cells with different temperatures and velocities. Therefore, absorption lines can have different Doppler shifts compared to the same lines in emission. When the light rays from all regions of the planet's visible side are combined, the central wavelengths of the absorption and emission features do not necessarily coincide, leading to complex line shapes set by a superposition of both emission and absorption features.

For example, consider the complex shape of the Doppler-shifted $\sim$2.30928 $\mu$m H$_2$O feature in the active cloud spectrum at $\varphi=225^\circ$ (Figure \ref{fig:spectra}). This feature appears to have both a blueshifted absorption component and a redshifted emission component. At this orbital phase, much of the temperature-inverted region (at higher dayside latitudes) has crossed the sub-observer longitude and is receding away from the observer. Additionally, on the western limb of the planet, a non-thermally-inverted region is approaching the observer. The combined effect of this geometry is a spectrum that contains a redshifted emission line from the receding portion of the atmosphere and a blueshifted absorption line from the approaching side of the planet. 

This Doppler splitting of spectral features is a unique characteristic of the active cloud GCM because of the active clouds' ability to create inhomogeneous thermal inversions in the upper atmosphere. Post-processed cloud models lack these spectral features because the clouds do not directly influence the temperature structure in the GCM, and hence cannot create inhomogeneous thermal inversions. We continue our discussion of Doppler splitting in the following section.

\subsection{Doppler Signatures} \label{sec:results:doppler}

From Figure \ref{fig:spectra} (and Figures \ref{fig:all_spectra1} and \ref{fig:all_spectra2} in Appendix \ref{sec:app:figures}) we can see that Doppler effects strongly influence the emission spectrum line profiles, and that these effects are phase-dependent. The shape, depth, and position of each line are dependent on a complex combination of wind and rotation velocities along the line-of-sight, inhomogenous clouds that act as an additional source of opacity, and atmospheric temperatures that vary in latitude, longitude, and altitude. As discussed previously, spatially non-uniform temperature inversions (caused by clouds) can also cause spectral lines to appear simultaneously as emission and absorption features, but with different Doppler signatures.

\begin{figure*}[t]
    \centering
    \includegraphics[width=\textwidth]{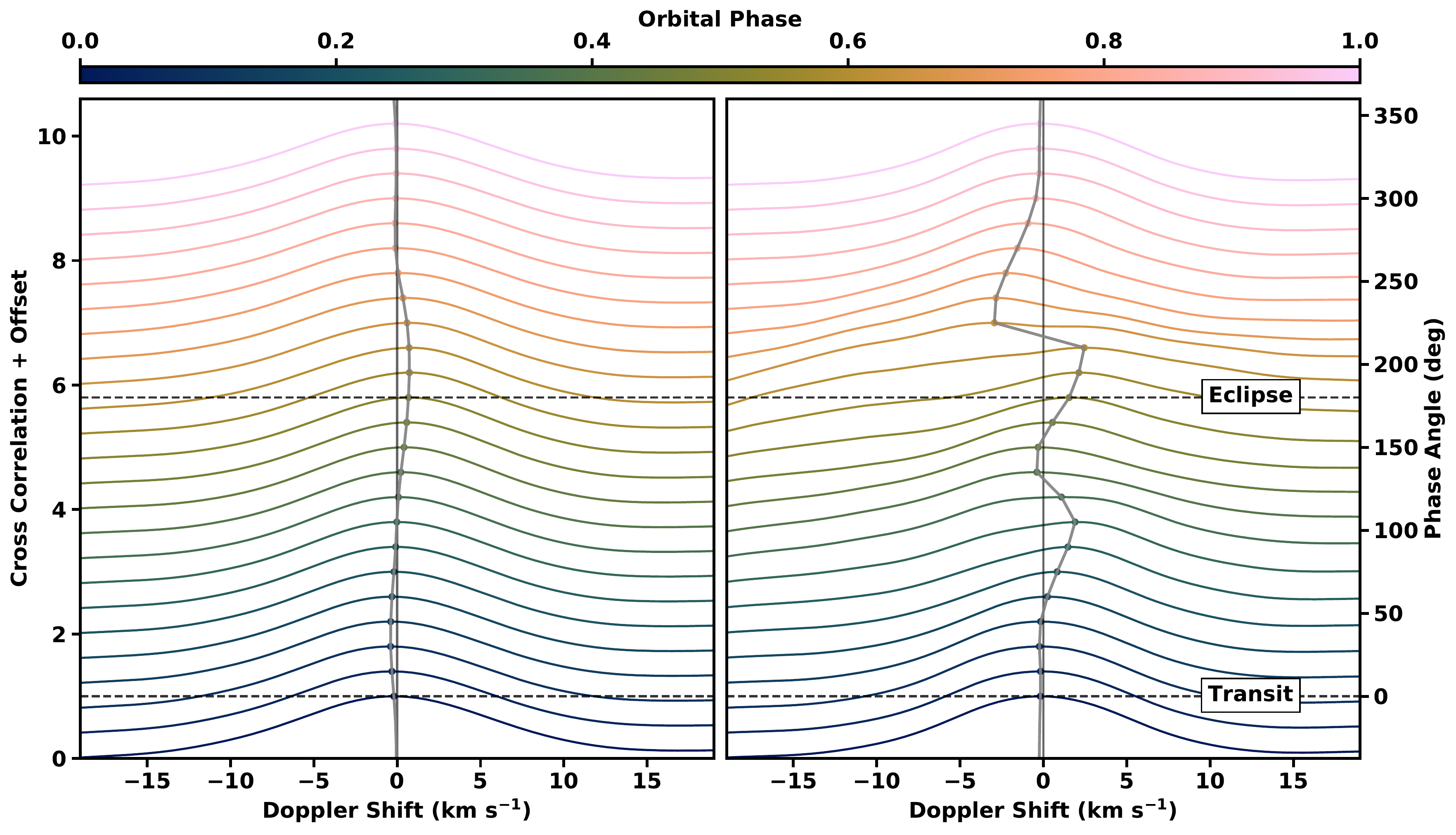}
    \begin{minipage}[]{0.49\textwidth}
        \centering
        (a) Clear
    \end{minipage}
    \begin{minipage}[]{0.49\textwidth}
        \centering
        (b) Active Clouds
    \end{minipage}
    \caption{Cross-correlation between Doppler-shifted and corresponding unshifted spectra shown as a function of orbital phase (indicated by the color scale). Each cross-correlation function (CCF) has been scaled and shifted by a constant offset such that the peak of each curve is aligned with the phase angle for that spectrum. The Doppler shifts of the CCF peaks are shown as a function of orbital phase (right-hand side vertical axis) by the solid gray line. Panel (a) shows CCFs resulting from the spectra of the cloud-free model and panel (b) shows the results from the active cloud model with extended thick clouds. The broadening and shifting of the CCFs reflects the Doppler broadening and shifting of resolved spectral lines. This arises from our geometrically self-consistent radiative transfer through the rotating planet's 3D wind and temperature field. Note the prevalence of complex (and for some phases, bimodal) CCFs in the active cloud case. At orbital phases where the dominant mode switches from the blueshifted side to the redshifted side of the CCF (e.g., $\varphi \approx 225^\circ$), there can be sharp discontinuities in the net Doppler shift function. See Figure \ref{fig:doppler_comparison} to compare the net Doppler shift functions for our other models.}
    \label{fig:ccfs}
\end{figure*}

\begin{figure*}[t]
    \centering
    \includegraphics[width=\textwidth]{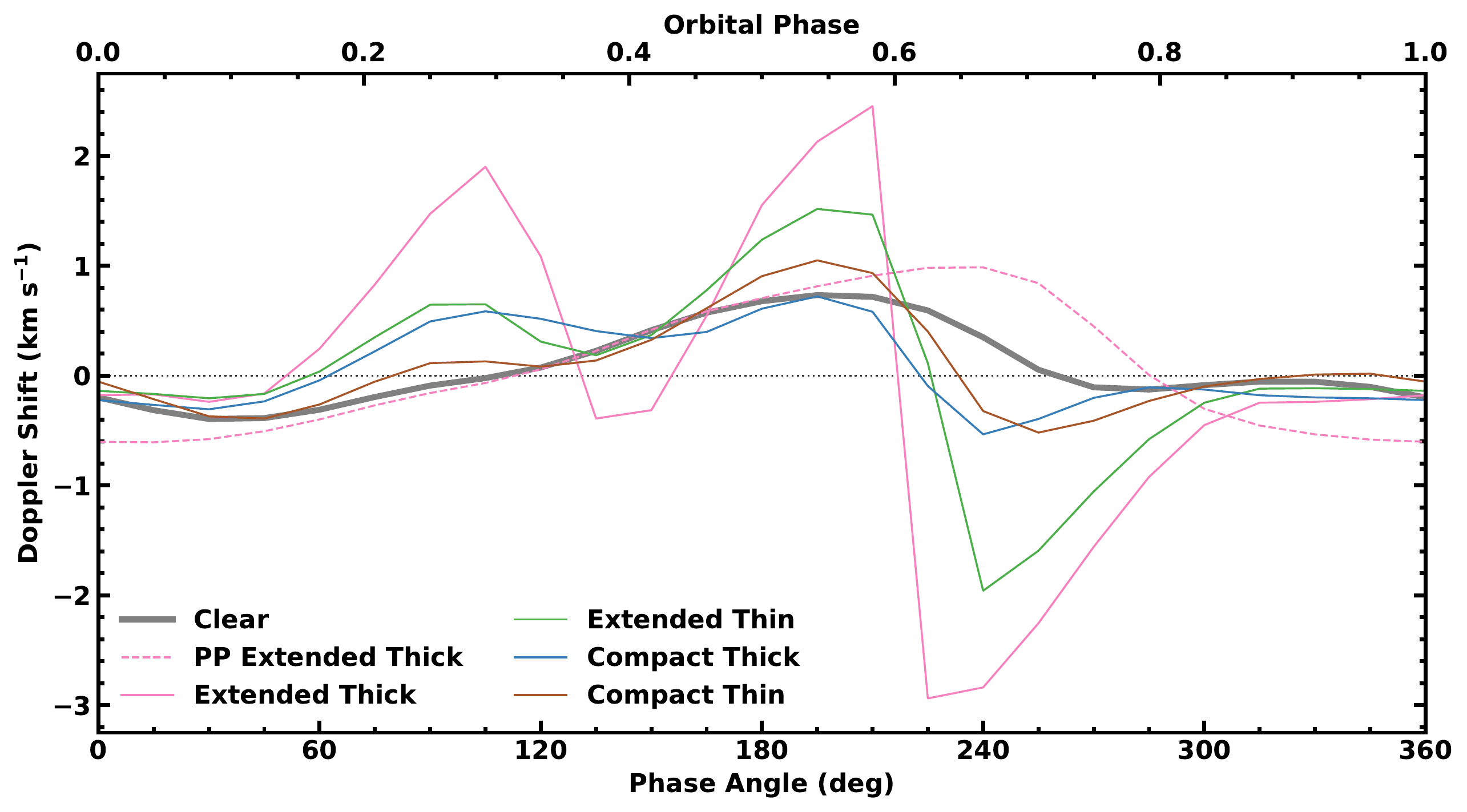}
    \caption{Net Doppler shift functions calculated from the cross-correlation of computed spectra with and without atmospheric Doppler-shifts (see Figure \ref{fig:ccfs}). Doppler shifts from the cloud-free model are shown by the solid gray line along with various active cloud GCM cases and the post-processed extended thick cloud case. As we transition from models whose clouds are more compact and thin to models whose clouds are more extended and thick, the Doppler shift functions differ more significantly from the cloud-free case and increase in amplitude. Doppler shifts from the post-processed model appear to be most similar to the Doppler shifts from the clear atmosphere (though not plotted here, all of the models with post-processed clouds have Doppler signatures that are qualitatively, and even quantitatively, similar to the clear model).}
    \label{fig:doppler_comparison}
\end{figure*}

\begin{figure}[t]
    \centering
    \includegraphics[width=0.49\textwidth]{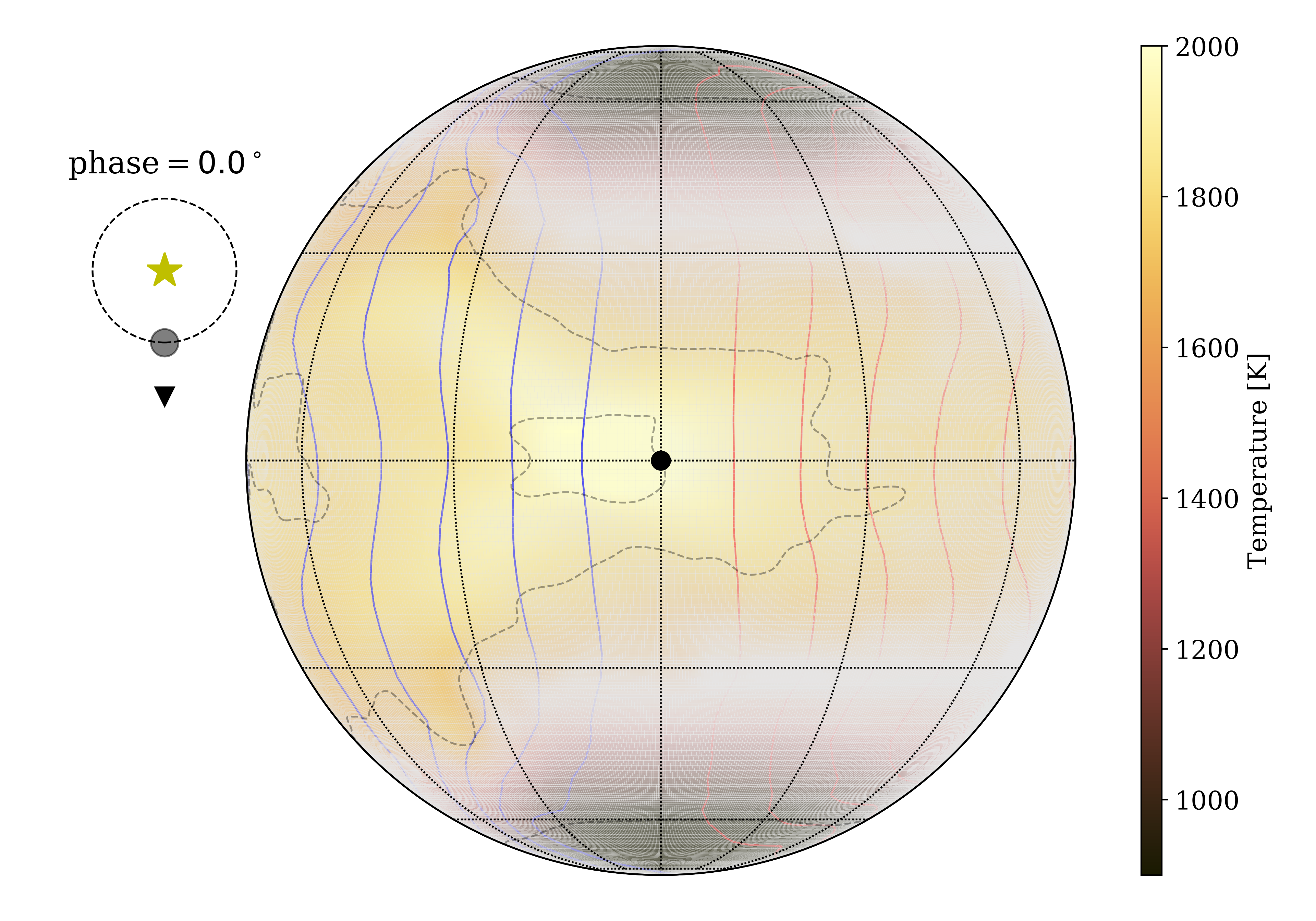}
    \caption{Temperature map of the planet's nightside ($\varphi=0^\circ$) from the active cloud GCM with extended thick clouds. Here, the vertical level in the atmosphere corresponds approximately to the infrared photosphere pressure level ($\sim$26 mbar) and the visible longitudes range from $90^\circ$ to $270^\circ$. The integrated optical depth of clouds above the photosphere is shown by the gray shaded area, with more opaque regions corresponding to thicker clouds. The red (blue) contours show the net positive (negative) line-of-sight velocities (in increments of 1 km\,s$^{-1}$) caused by atmospheric winds and the planet's rotation at the photosphere level, and the dot in the center of the planet indicates the anti-stellar point. The diagram on the left-hand side shows the location of the planet in its orbit as seen from the perspective of a distant observer, indicated by the triangle. This is representative of the animation available in the online journal, which loops through one orbital period of the planet, showing how the visible temperature pattern changes throughout the orbit.}
    \label{fig:movie_frame}
\end{figure}

To quantify how the sum of these factors affects the overall Doppler shift of the disk-integrated spectra as a function of orbital phase, we performed the following analysis. We determined the net Doppler shift at each orbital phase from our computed emission spectra by cross-correlating the Doppler-shifted spectrum with the corresponding unshifted template spectrum (calculated with Doppler effects turned off in the radiative transfer routine). We took the net Doppler shift at each phase to be the velocity shift corresponding to the peak of the cross-correlation function (CCF). The normalized CCFs from the clear and active cloud GCM spectra are shown for 24 orbital phases in Figure \ref{fig:ccfs}. The net Doppler shift as a function of orbital phase is indicated by the line connecting the CCF peaks in Figure \ref{fig:ccfs} and this is shown for all the active cloud cases, along with the clear case and the post-processed extended/thick cloud case in Figure \ref{fig:doppler_comparison}.

For the clear atmosphere, the cross-correlation functions are well-behaved and we recovered Doppler shifts similar to those found by \citet{zhang+2017} for cloudless 3D GCMs representative of HD~209458b, WASP-69b, and HD~189733b. Broadly speaking, the net Doppler shifts are dominated by the relative motion of the planet's brightest region (i.e., the hotspot to the east of the substellar longitude). Net Doppler blueshifts were produced when the brightest region of the planet was approaching the observer and net Doppler redshifts were produced when the brightest region was receding.

Because the brightest region of the planet is advected to a longitude approximately 45 degrees east of the substellar point, it rotates into view on the approaching limb of the planet when the orbital phase is about 45 degrees past transit ($\varphi \approx 45^\circ$). Emission from this region dominates over dimmer emission from the receding limb, creating the maximum net blueshift of $\sim$0.5 km\,s$^{-1}$ we see at $\varphi \approx 45^\circ$ in Figures \ref{fig:ccfs} and \ref{fig:doppler_comparison}. Around the time when the bright spot reaches the sub-observer longitude ($\varphi \approx 135^\circ$), the Doppler shift transitions from a net blueshift to a net redshift as the bright spot begins moving away from the observer. The brightest region then reaches the receding limb of the planet at an orbital phase of $\varphi \approx 225^\circ$, lagging secondary eclipse by approximately 45 degrees, creating the maximum redshift of $\sim$1 km\,s$^{-1}$ at $\varphi \approx 225^\circ$. Smaller-scale variations in the Doppler shift pattern are more difficult to explain qualitatively because they result from the combined effects of various properties of the complex 3D atmosphere, as mentioned above.

Spectra from our model with post-processed clouds had the most similar Doppler shifts to those from the clear atmosphere. This result is reasonable because the temperatures and wind structure of these two models was identical. Differences in the Doppler signatures therefore arose purely from the excess gray opacity from clouds. Since the cloud coverage was non-uniform, the deviations of the Doppler shifts from the clear model depend on the position of optically thin regions in the cloud distribution and hence are phase-dependent. Consider, for example, the Doppler shifts around orbital phase $\varphi \approx 345^\circ$ in Figure \ref{fig:doppler_comparison} (indicated by the dashed pink line for the post-processed cloud model). Here we see an enhanced blueshift relative to the clear model, which can be explained by fact that clouds on the planet's dayside are generally thinner than the clouds on the nightside, especially at longitudes east of the substellar point (see Figure \ref{fig:temp_maps}). At $\varphi = 345^\circ$, the visible longitudes range from $+105^\circ$ to $-75^\circ$; thus, the approaching limb of the planet is relatively clear of clouds, while the receding limb is almost fully concealed by clouds. This attenuates the emission from the redshifted side of the planet, resulting in a stronger net blueshift. The inverse of this effect can be seen at an orbital phase of $\varphi \approx 255^\circ$, when the approaching limb of the planet is covered by clouds and the receding limb is relatively cloud-free. 

It is obvious by looking at Figures \ref{fig:ccfs} and \ref{fig:doppler_comparison} that the Doppler shifts from the active cloud model are much more complex and variable than those from the clear (or post-processed) model. The active cloud cross-correlation functions do not have the same quasi-Gaussian shape as the CCFs for the clear model and they vary in a more unexpected manner as a function of phase. An exact diagnosis of the shapes of the CCFs is challenging because the Doppler signals come from a mixture of spectral lines that originate from different spatial locations on the planet (and different heights in the atmosphere), and are each affected by a complex combination of rotation speed and inhomogeneous winds, temperatures, and cloud opacities. 

We can attempt to understand the general behavior of the net Doppler shifts by tracking the relative motion of the regions of the atmosphere with temperature inversions (located around the higher dayside latitudes and along the meridian at a longitude of about $-90$ degrees; see Figure \ref{fig:tp+inversion}) and the planet's brightest region. To aid with our interpretation of the Doppler shifts, we utilized an animation showing the visible temperature structure of the planet throughout the planet's orbit to evaluate when the planet's brightest region and thermally inverted regions were either approaching or receding away from the observer. A representative snapshot from the animation of the active cloud atmosphere with extended thick cloud properties is shown in Figure \ref{fig:movie_frame}. Note that the brightest region of the active cloud atmosphere is shifted east of the substellar longitude by about 30 degrees, less than it was in the clear atmosphere \citep[as noted by][]{roman+2019}.

At $\varphi \approx 0^\circ$ (i.e., transit geometry), both the bright spot and the thermally inverted regions are hidden from view. The visible temperature pattern is relatively uniform, so this alignment produces a spectrum similar to one from the clear atmosphere, as well as a simple CCF and almost negligible net Doppler shift. As the bright spot starts to become visible on the approaching limb, the spectra become slightly blueshifted. However, this blueshift is weaker than the clear model because emission features from the thermally inverted higher latitudes effectively decrease the strength of absorption features on the approaching limb relative to those on the receding limb.

As more regions with temperature inversions became visible, the emission from these regions effectively dilutes the absorption features on the approaching limb enough that the absorption features on the receding limb began to dominate the net signal. This leads to a gradual transition from a net blueshift to a net redshift, which occurs at an orbital phase of $\varphi \approx 60^\circ$, followed by a steady increase in the amplitude of the net redshift up to $\sim$2 km\,s$^{-1}$ at $\varphi \approx 105^\circ$.

The planet's brightest region then approaches and crosses over the sub-observer longitude, at an orbital phase of $\varphi \approx 120^\circ$. At this point, the distribution of visible high-latitude temperature inversions becomes approximately symmetric (the inversions along the western terminator are still hidden from view). Hence the net Doppler shift of the spectra is approximately zero.

Note that the high-altitude thermal inversions can sometimes produce emission reversals near to the core of certain strong spectral lines (e.g., the $\sim$2.30892 $\mu$m feature in the active cloud Doppler-shifted spectra in Figure \ref{fig:all_spectra1} [Appendix \ref{sec:app:figures}] has a small perturbation in the absorption line that moves from shorter wavelengths to longer wavelengths as phase increases from $\varphi \approx 90^\circ$ to $\varphi \approx 135^\circ$). This effect is akin to the more familiar emission reversals in the cores of the CaII H \& K lines in the spectra of main-sequence stars, although in our case the line shapes are more complex due to the asymmetry in the temperature field caused by the planet being irradiated from one side only. In our spectra, this effect can lead to the complex (sometimes bimodal) appearance of the CCFs in Figure \ref{fig:ccfs}.

Because the net Doppler shift is determined only by the peak of the CCF, which corresponds the to wavelength associated with the \emph{deepest} part of the lines, it is sensitive to the relative positions of both the brightest regions and the temperature inversions. This explains why there is a transition from redshift to blueshift at $\varphi \approx 120^\circ$ in Figure \ref{fig:doppler_comparison}. The sensitivity of the net Doppler shift signal demonstrates a limitation of interpreting observed spectra using 1D models, which cannot capture the complexity of the spectra that arises from inherently 3D effects.

Next, there is a net blueshift as absorption features originating from the approaching limb dominate over the signal from the receding limb. This gradually transitions back to a net redshift, reaching a net zero Doppler shift at a phase of $\varphi \approx 150^\circ$ and a maximum redshift of $\sim$2.5 km\,s$^{-1}$ at $\varphi \approx 210^\circ$. At $\varphi \approx 180^\circ$, the temperature inversions along the western terminator start to become visible on the approaching side of the planet, creating strong blueshifted emission lines. As before, these emission lines create perturbations in the absorption features, which gradually transition from shorter wavelengths to longer wavelengths as phase increases (Figure \ref{fig:all_spectra1}). Again, this leads to spectral lines with two local minima, giving rise to complex CCF shapes. This explains the bimodal CCF and sharp transition from the maximum redshift to the maximum blueshift ($\sim$3 km\,s$^{-1}$) at $\varphi \approx 225^\circ$ (Figures \ref{fig:ccfs} and \ref{fig:doppler_comparison}).

The Doppler shift functions for the other active cloud models in Figure \ref{fig:doppler_comparison} lie somewhere between the behavior for the atmosphere with extended thick clouds and the clear atmosphere. These can be explained in a similar way to the extended thick cloud case, since all of the active cloud GCMs exhibit stronger temperature inversions than the clear atmosphere (see Figure \ref{fig:all_tp+inversion} in Appendix \ref{sec:app:figures}). As expected, there are increasingly strong temperature inversions as we transition from GCMs with thinner and more compact clouds to GCMs with thicker and more extended clouds. Likewise, the Doppler shift functions become more distinct from the clear model (with greater shifts and sharper discontinuities) as the temperature inversions in the GCMs become more significant.

We note that while Doppler \emph{broadening} is another interesting feature of the spectra, extracting information from it is not straightforward. \citet{zhang+2017} showed that the broadening has significant contributions from both rotation and winds (which are largely in the same direction as the rotation), with the relative contributions depending on the particular planet. An important consequence of this is that we cannot strongly constrain the rotation rate because it does not uniquely define the broadening \citep{beltz+2020}. Moverover, the broadening in the cloudy versions of the spectra is complex because the shapes of the CCFs are superpositions of multiple components. Therefore, we do not include a detailed interpretation of Doppler broadening here because we are unable to provide an unambiguous interpretation.

%
%
\section{Conclusions} \label{sec:conclusion}

We have presented radiative transfer calculations of high-resolution thermal emission spectra from 3D models of hot Jupiter atmospheres with radiatively active clouds. This is the first time for hot Jupiters that a radiative transfer emission spectrum calculation from a 3D GCM has accounted for line-of-sight geometry with non-uniform aerosol extinction, as well as 3D thermal structure, global winds, and rotation velocities. We have demonstrated that emission spectra are highly sensitive to clouds at high resolution, due to their radiative effect on the atmosphere. 

Compared to a clear atmosphere, cloudy atmosphere spectra show a number of fundamental differences---both in the continuum levels and line shapes. On the other hand, when post-processed (passive) clouds are added to an otherwise clear atmosphere, there is a minimal effect on the spectra. This is because the thermal structure of the atmosphere is significantly altered by the presence of clouds \citep{roman+2019}. We recovered $\sim$km/s net Doppler shifts from our spectra by cross-correlating them with spectra calculated from the same atmospheric models but without atmospheric Doppler effects included, and found that the thermal differences in the cloudy atmosphere led to spectral Doppler signatures that were markedly different from those in the clear and post-processed cloud models. 

Our conclusions are summarized as follows:

\begin{itemize}
    \item Despite the clouds having the same optical properties and vertical extent in the active and post-processed cloudy models, the radiative feedback between the active clouds and the thermal structure of the atmosphere subsequently leads to drastically different thermal emission spectra.
    \item Spectra from the active cloud atmosphere generally have weaker absorption features and the dayside spectra are especially flattened because of the presence of thermal inversions induced by clouds. Some emission features appear in partial superposition with absorption features in the spectra when regions of the planet with thermal inversions have a different net Doppler shift than regions producing absorption lines.
    \item The continuum flux from the active cloud model is generally weaker than the continuum flux from the clear and post-processed cases, and the contrast between the brightest and dimmest continuum levels is weaker, in agreement with the bolometric results in \citet{roman+2019}.
    \item Thicker and more extended clouds result in emission spectra that differ the most substantially from a clear atmosphere. Since our GCM does not self-consistently include sophisticated treatments of clouds, we cannot comment on the expected vertical extent or total optical thickness of the clouds in planets similar to Kepler-7b; we only conclude that thicker and more extended clouds produce more extreme effects on the atmosphere and subsequent high-resolution emission spectra.
    \item Net Doppler shifts recovered from the clear and post-processed cloud models are dominated by the motion of the planet's brightest region, with differences between the two models arising only from non-uniform cloud opacities.
    \item The presence of temperature inversions in the active cloud model make the Doppler shifts significantly more complex. Emission lines from thermally inverted regions of the planet lead to complex (sometimes bimodal) cross-correlation functions as they rotate into and out of view. This makes the peak cross-correlation value very sensitive to small changes viewing geometry, and subsequently leads to sharp changes in the net Doppler shift as a function of orbital phase.
\end{itemize}

We remind the reader that we have neglected the effects of non-isotropic scattering on our high-resolution spectra and performed our radiative transfer calculations in the limit of pure thermal emission. A more realistic treatment of scattering may be important over a larger wavelength range, especially where clouds are optically thick and have high single-scattering albedos and strongly non-isotropic phase functions. We intend to upgrade our spectral radiative transfer routine to a two-stream scattering approximation in future work.

We also note that our double-gray prescription for radiative transfer in the GCM does not capture the radiative feedback from the clouds in a fully self-consistent, multi-wavelength framework and ignores cloud microphysics. As such, and as discussed in more detail in \citet{roman+2019}, our cloud prescription (especially for the thick extended cloud realization) can be considered perhaps an over-estimate of the role of cloud feedback in influencing the planet's atmospheric state. This means that reality may be somewhere between our model with the strongest cloud effects and a clear or post-processed case. We also point out that, while our models lack a detailed treatment of cloud microphysics, we find that the strongest impact on our modeled high resolution spectra arise from the spatial distribution of the clouds and their impact on the thermal structure of the atmosphere. As a result, it may be that the details of the cloud particle properties have only a second-order effect.

Moreover, because our spectra were calculated from the final time step of the GCM, we did not assess the possible effects of temporal atmospheric variations on high-resolution emission spectra. Temporal variability in hot Jupiter atmospheres may be an important consideration for secondary eclipse and phase curve observations \citep{komacek+2020}. Hence, in future studies it is worth investigating how temporal variability may affect cloud distributions and subsequent high-resolution emission spectra and Doppler shifts in hot Jupiter atmospheres.

Finally, we reiterate the observational relevance of this study. From the very first use of HRS for exoplanet characterization \citep{snellen+2010}, there have been hints at this technique's ability to constrain atmospheric motions of hot Jupiters, with more stringent empirical constraints in subsequent work \citep{louden+2015, brogi+2016, flowers+2019}. In addition, \citet{zhang+2017} demonstrated the impact of a planet's three-dimensional temperature structure on these Doppler signatures in emission spectra, and other studies have analyzed the influence that aerosols can have on high-resolution spectra \citep{pino+2018, hood+2020, gandhi+2020}.

Especially as we look forward to the era of high-resolution spectroscopy with Extremely Large Telescopes (e.g., with planned instruments such as G-CLEF \citep{szentgyorgyi+2012} and GMTNIRS \citep{lee+2010} on the Giant Magellan Telescope, MODHIS \citep{mawet+2019} and MICHI \citep{packham+2012} on the Thirty Meter Telescope, and METIS \citep{Brandl2016} and HIRES \citep{zerbi+2014} on the European Extremely Large Telescope), the signatures of non-uniform 3-D atmospheric structure will be apparent in observed exoplanet emission spectra. In this expanding landscape, 3-D atmospheric models will be called upon and will need to include clouds in a more self-consistent manner to be able to accurately interpret their signatures in HRS data. We have shown that more simplistic treatments such as  post-processing clouds in atmospheric models may fail to predict the complex spectral characteristics that arise due to the inherent radiative feedback between clouds and the atmospheric thermal structure.

%
%
\acknowledgments

We thank the anonymous referee for their constructive comments which helped to improve this manuscript. This research was supported by NASA Astrophysics Theory Program grant NNX17AG25G. C.K.H. acknowledges support from the National Science Foundation Graduate Research Fellowship Program under Grant No. DGE1752814, and additional support from the University of Maryland Astronomy Department Honors Program.

\software{numpy\,\citep{numpy},\,matplotlib\,\citep{matplotlib}}

\bibliography{ms}

\appendix

%
%
\section{Additional Figures} \label{sec:app:figures}

\begin{figure*}[h]
    \centering
    \begin{minipage}[]{0.4\textwidth}
        \centering
        (a) Extended Thick
        \includegraphics[width=\textwidth]{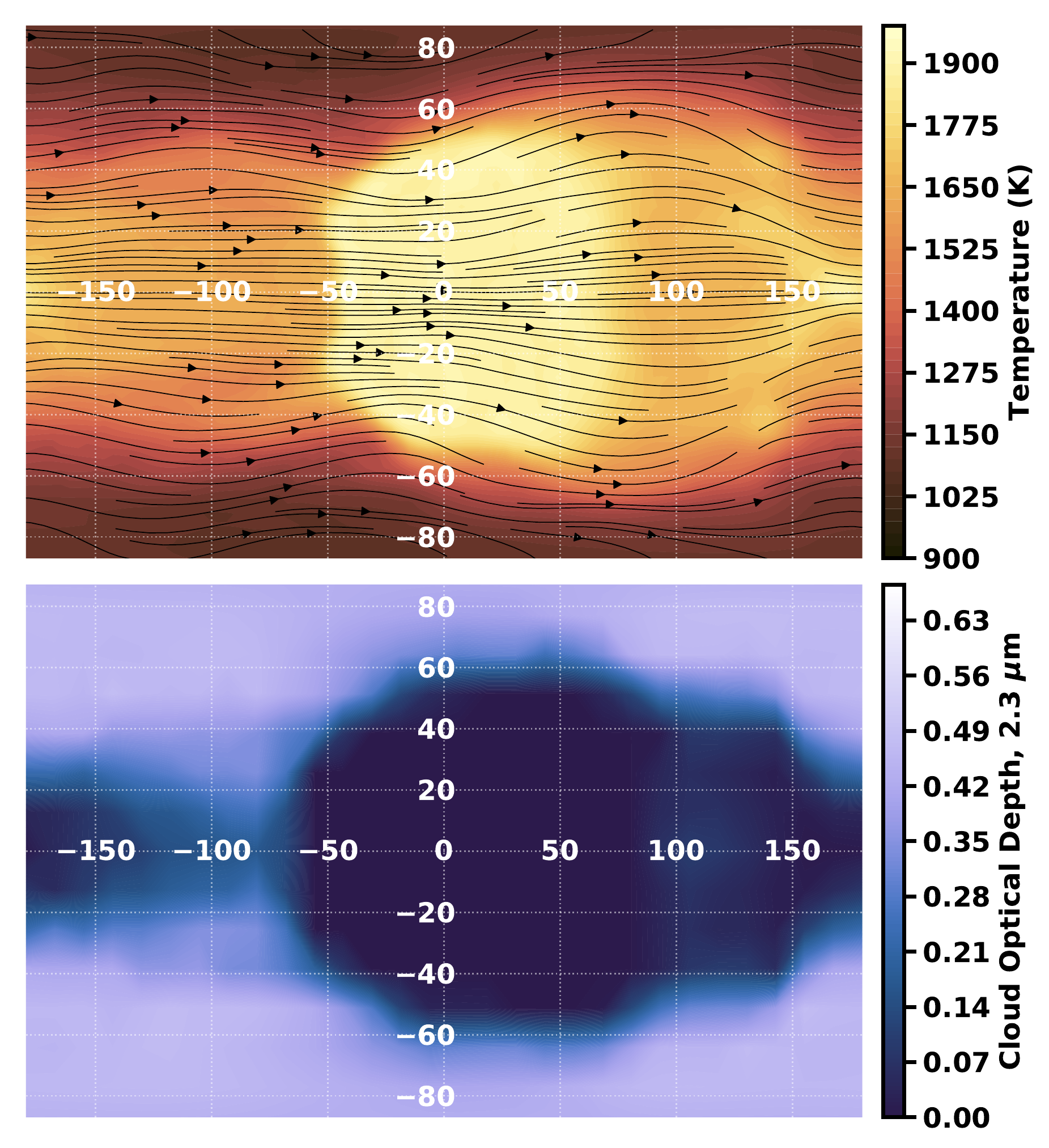}
    \end{minipage}
    \begin{minipage}[]{0.4\textwidth}
        \centering
        (b) Compact Thick
        \includegraphics[width=\textwidth]{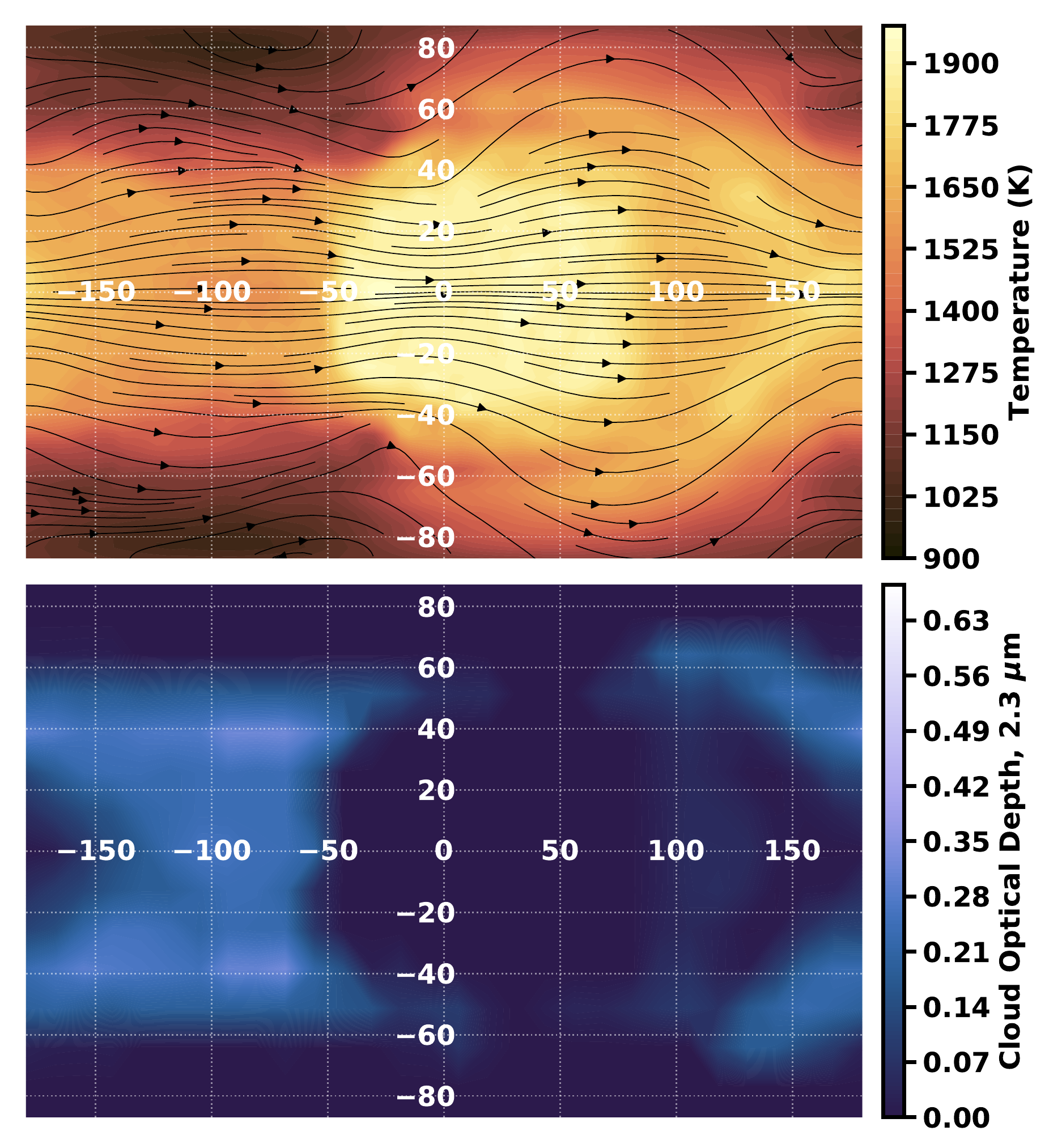}
    \end{minipage} \\
    \begin{minipage}[]{0.4\textwidth}
        \centering
        \includegraphics[width=\textwidth]{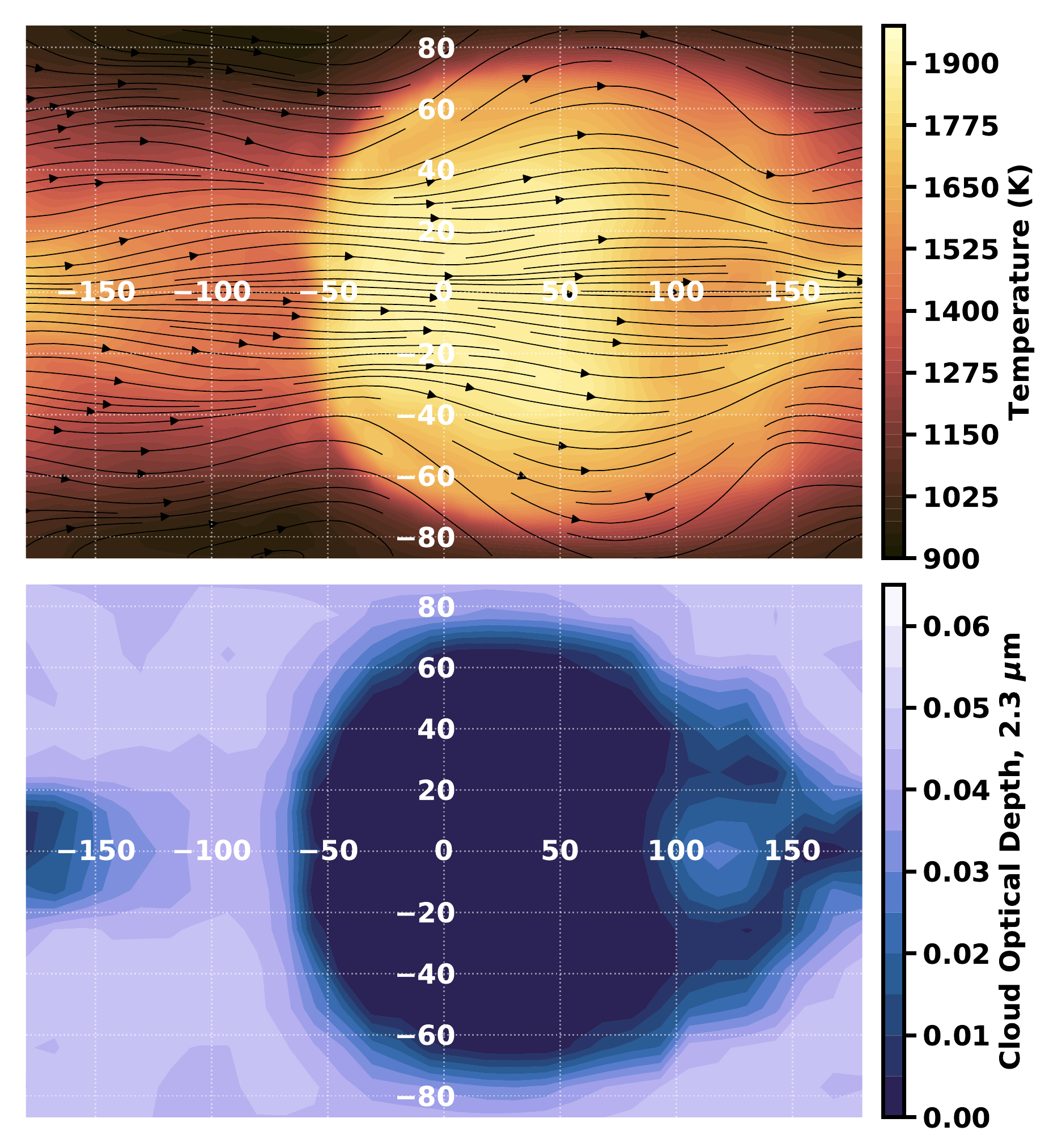}
        (c) Extended Thin
    \end{minipage}
    \begin{minipage}[]{0.4\textwidth}
        \centering
        \includegraphics[width=\textwidth]{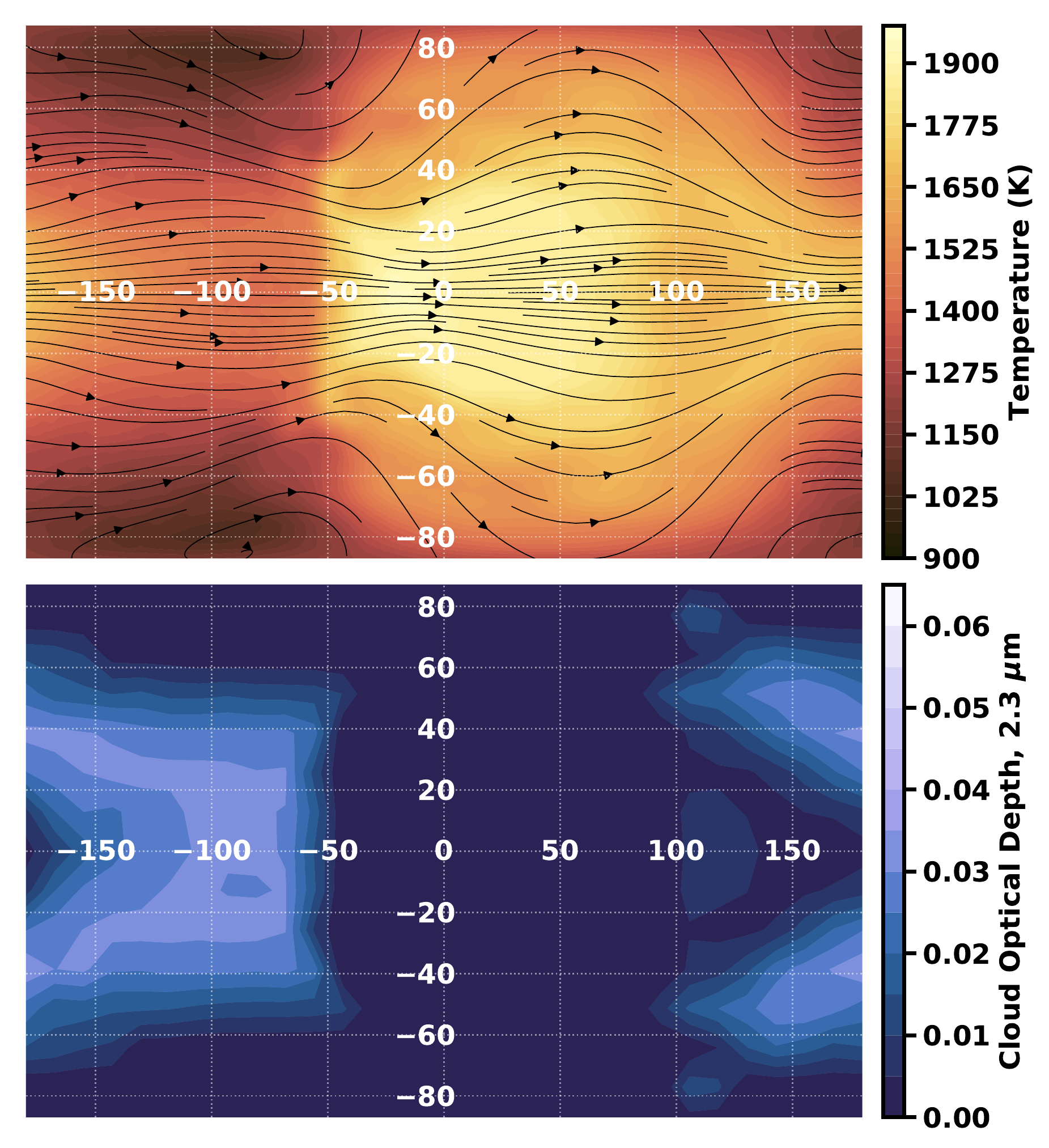}
        (d) Compact Thin
    \end{minipage}
    \caption{Longitude-latitude maps of simulated temperatures, winds, and infrared (2.3-micron) cloud optical depth from active cloud GCMs with extended/thick clouds (a), compact/thick clouds (b), extended/thin clouds (c), and compact/thin clouds (d) from \citet{roman+2019}. As in Figure \ref{fig:temp_maps}, the temperatures and wind vectors correspond approximately to the infrared photosphere pressure level of the clear model ($\sim$26 mbar) and the IR cloud optical depths are integrated vertically above the photosphere. Note that the values of optical depth on the colorbar for the thin cloud models are 10\% of the colorbar values for the thicker cloud models. The extended/thick cloud model has the most extensive and optically thick cloud coverage and a thermal structure least similar to that of the clear model. As we transition to models whose clouds are more compact and thin, the cloud coverage and thermal structure become increasingly similar to the cloud-free model.}
    \label{fig:all_temp_maps}
\end{figure*}

\begin{figure*}[h]
    \centering
    \begin{minipage}[]{0.45\textwidth}
        \centering
        (a) Extended Thick
        \includegraphics[width=\textwidth]{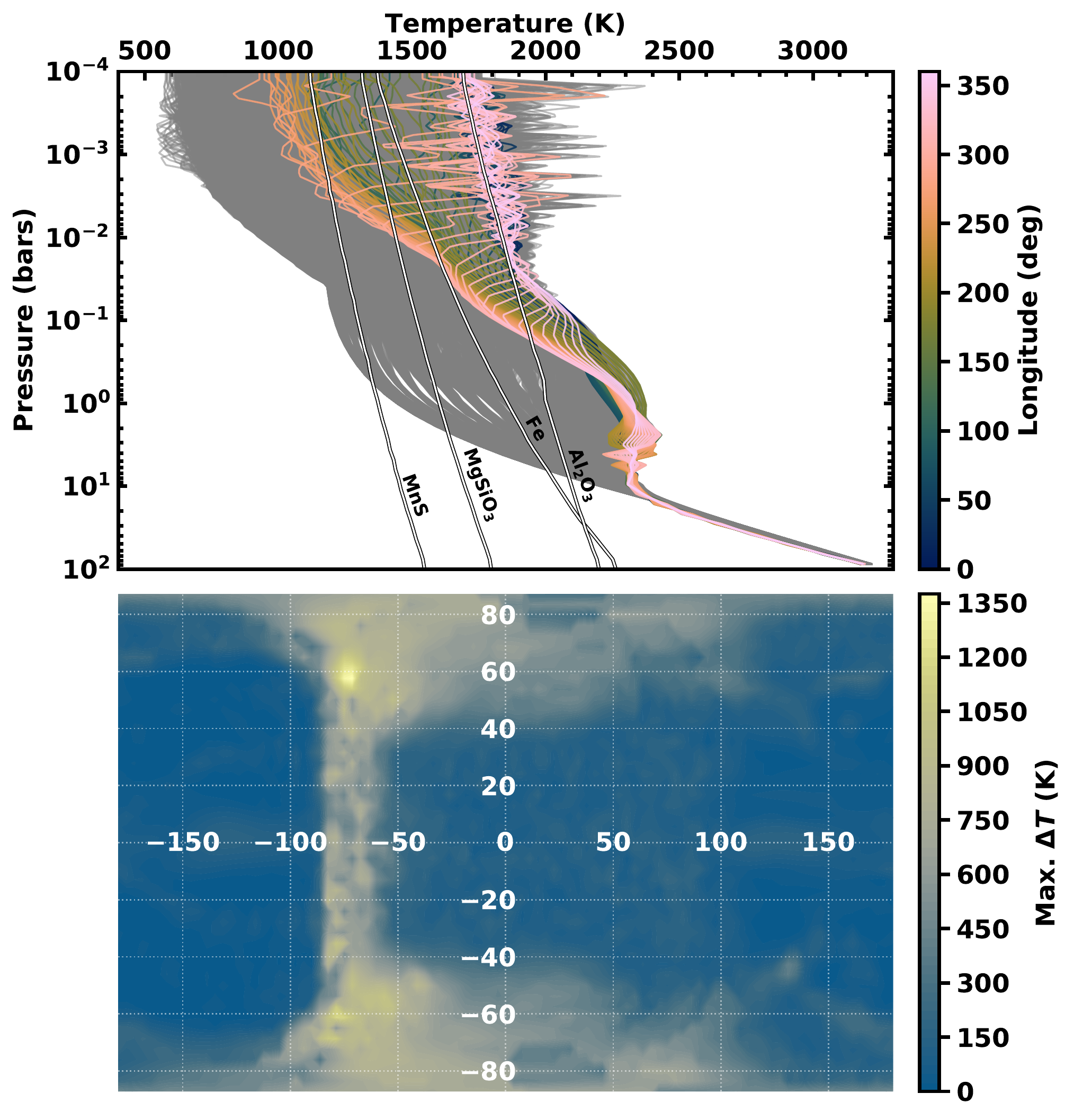}
    \end{minipage}
    \begin{minipage}[]{0.45\textwidth}
        \centering
        (b) Compact Thick
        \includegraphics[width=\textwidth]{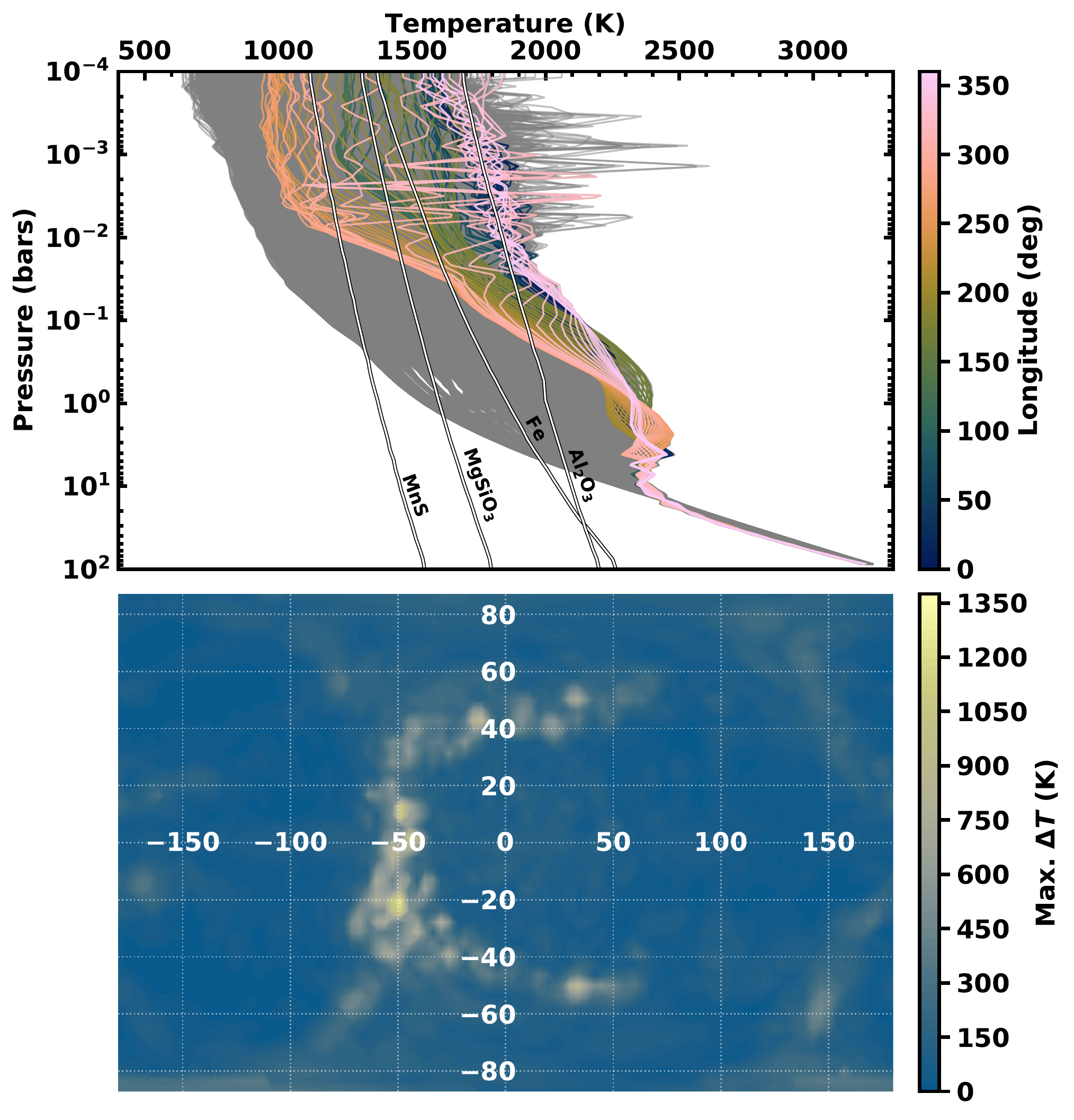}
    \end{minipage} \\
    \begin{minipage}[]{0.45\textwidth}
        \centering
        \includegraphics[width=\textwidth]{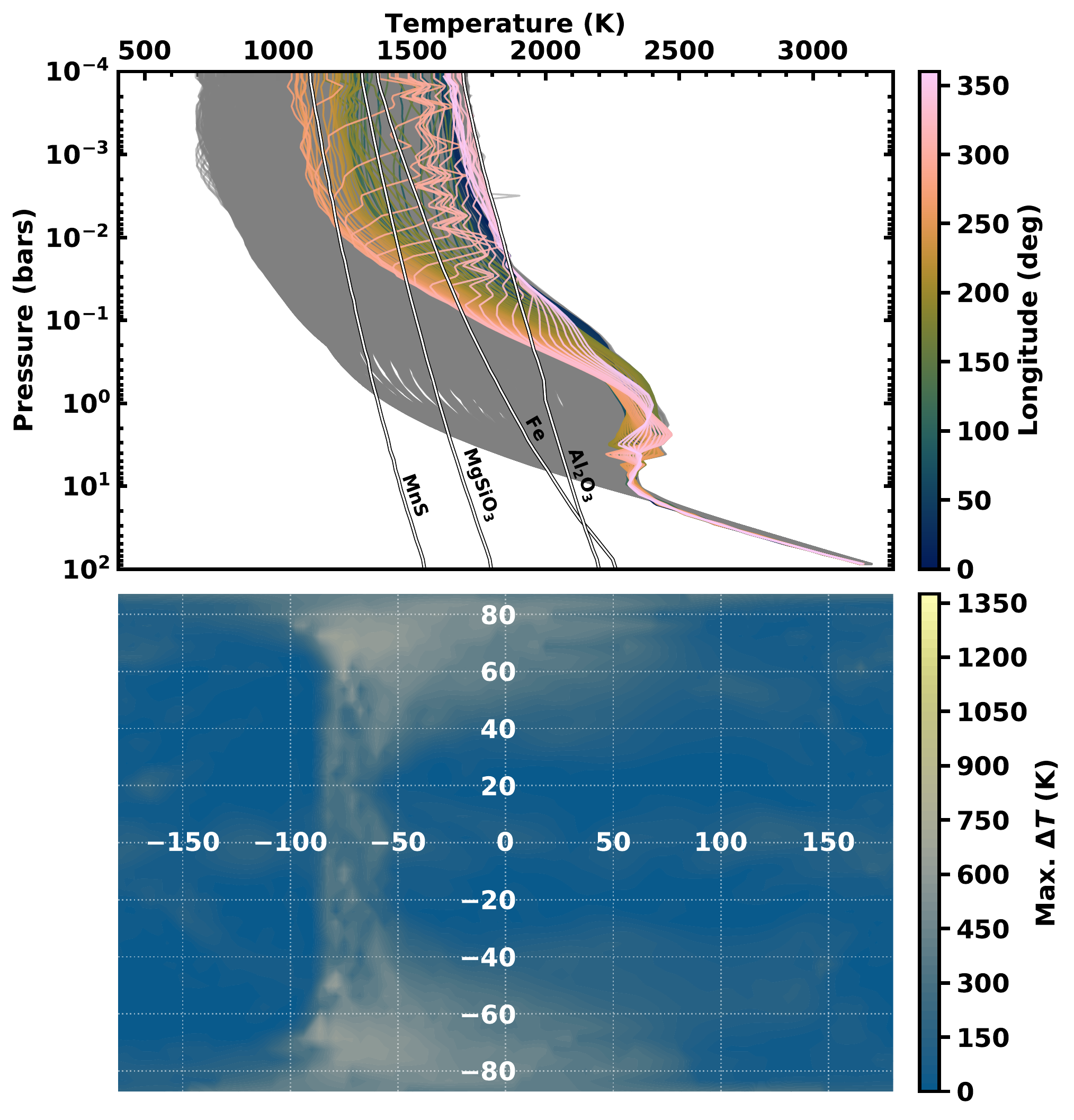}
        (c) Extended Thin
    \end{minipage}
    \begin{minipage}[]{0.45\textwidth}
        \centering
        \includegraphics[width=\textwidth]{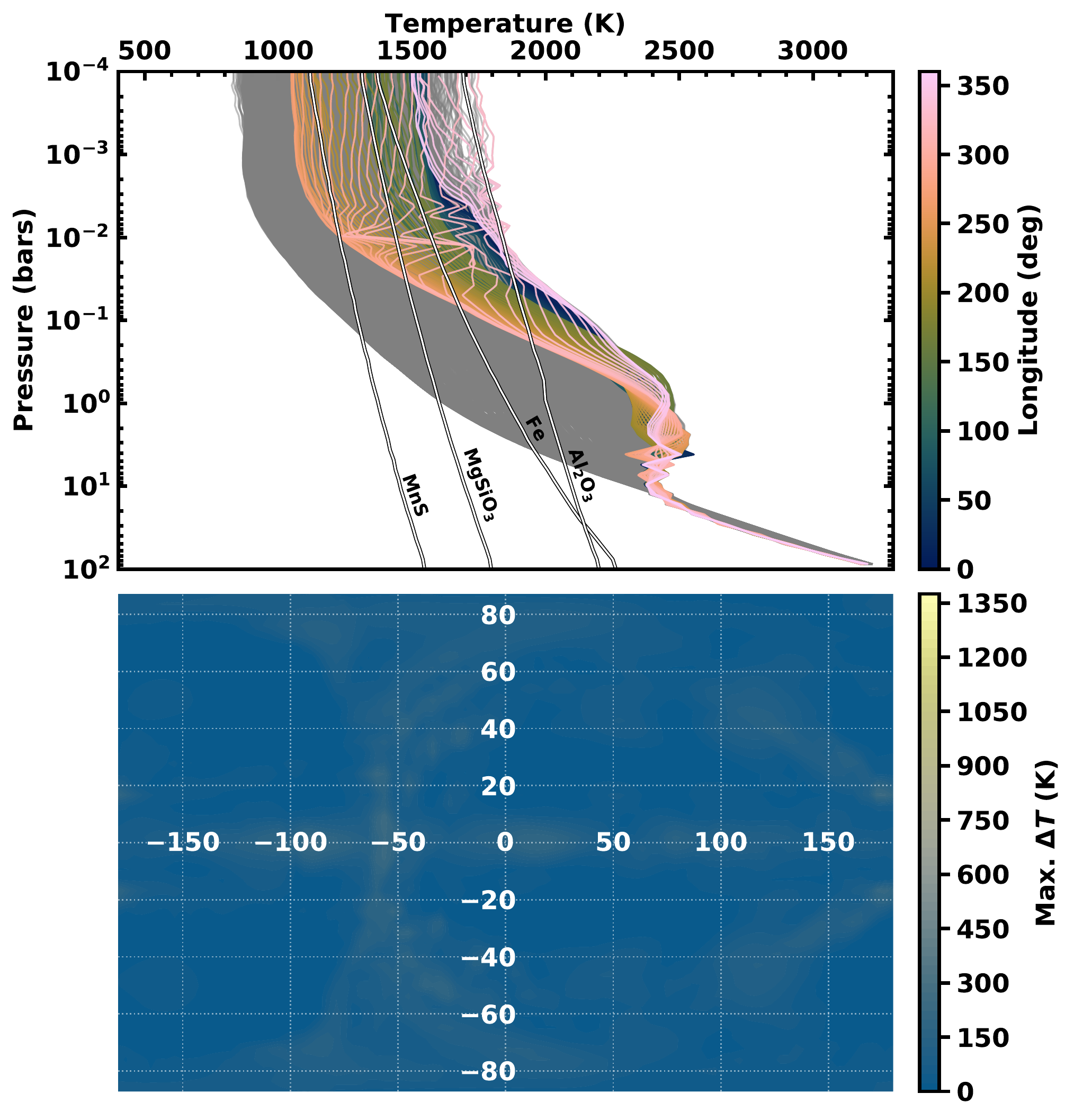}
        (d) Compact Thin
    \end{minipage}
    \caption{Vertical temperature profiles and latitude-longitude maps of maximum vertical temperature inversions from active cloud GCMs with extended/thick clouds (a), compact/thick clouds (b), extended/thin clouds (c), and compact/thin clouds (d). As in Figure \ref{fig:tp+inversion}, Cloud condensation curves from \citet{roman+2019} are labeled for each species. The color temperature profiles are located at the planet's equator at the longitude indicated by the color scale, while gray profiles are located at all other latitudes and longitudes. The strong temperature inversions along the western terminator, which are caused by clouds trapping heat, become more apparent as we transition from models with clouds that are more compact and thin to models with clouds that are more extended and thick.}
    \label{fig:all_tp+inversion}
\end{figure*}

\begin{figure*}[h]
    \centering
    \includegraphics[width=0.9\textwidth]{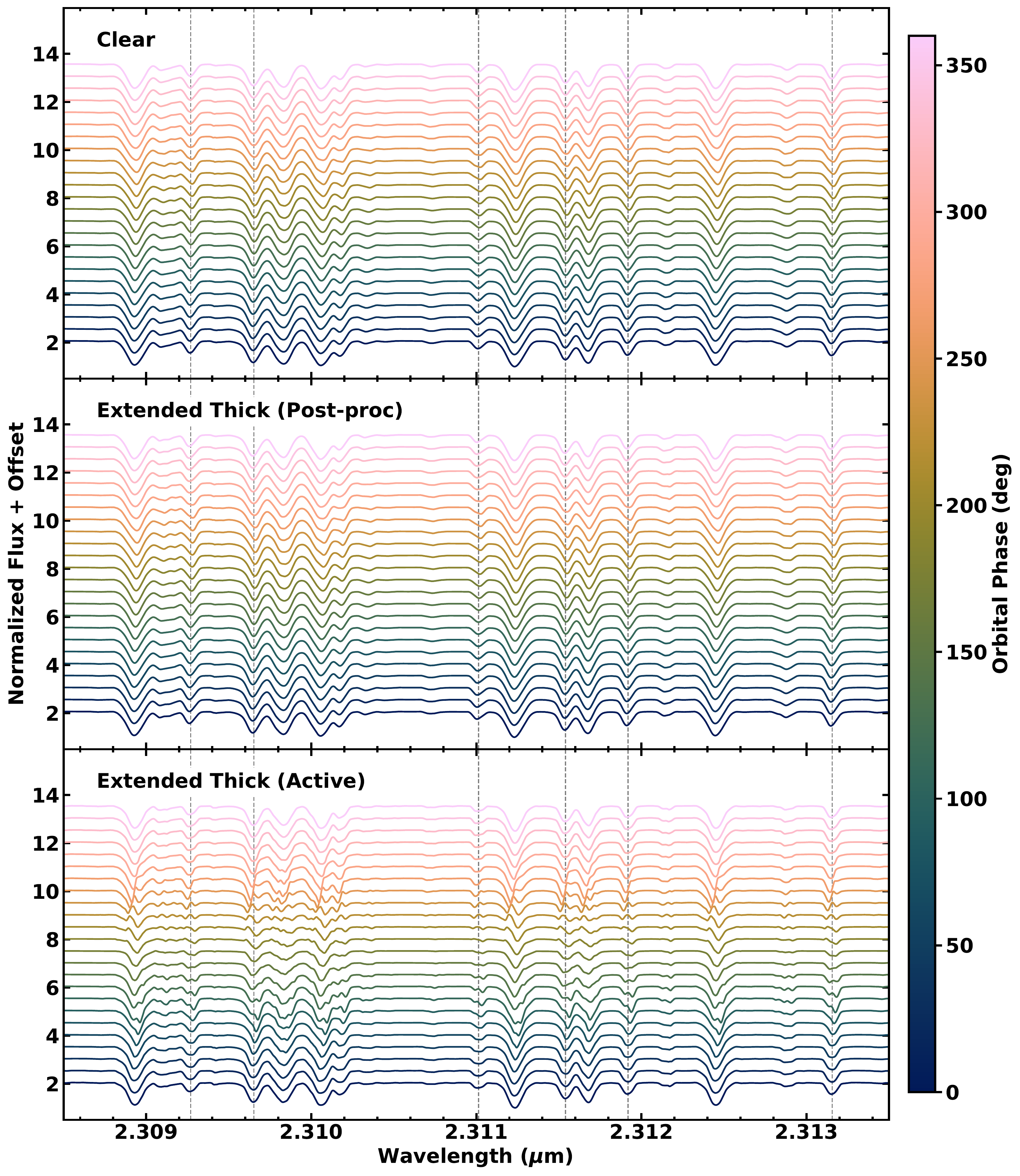}
    \caption{Doppler-shifted emission spectra for 24 orbital phases from the clear GCM (top), post-processed cloud GCM (middle), and active cloud GCM (bottom), shown as normalized flux with an arbitrary, constant offset. Both sets of cloudy spectra are calculated from the models with extended thick clouds. The color scale indicates the orbital phase, and hence which side of the planet is visible, used to calculate each spectrum. Dashed vertical lines indicate the rest-frame line centers of several prominent water features. Changes in the broadening and shifting due to Doppler motion are most apparent in the active cloud spectra, which is consistent with the net Doppler shift function shown in Figure \ref{fig:doppler_comparison}.}
    \label{fig:all_spectra1}
\end{figure*}

\begin{figure*}[h]
    \centering
    \includegraphics[width=0.9\textwidth]{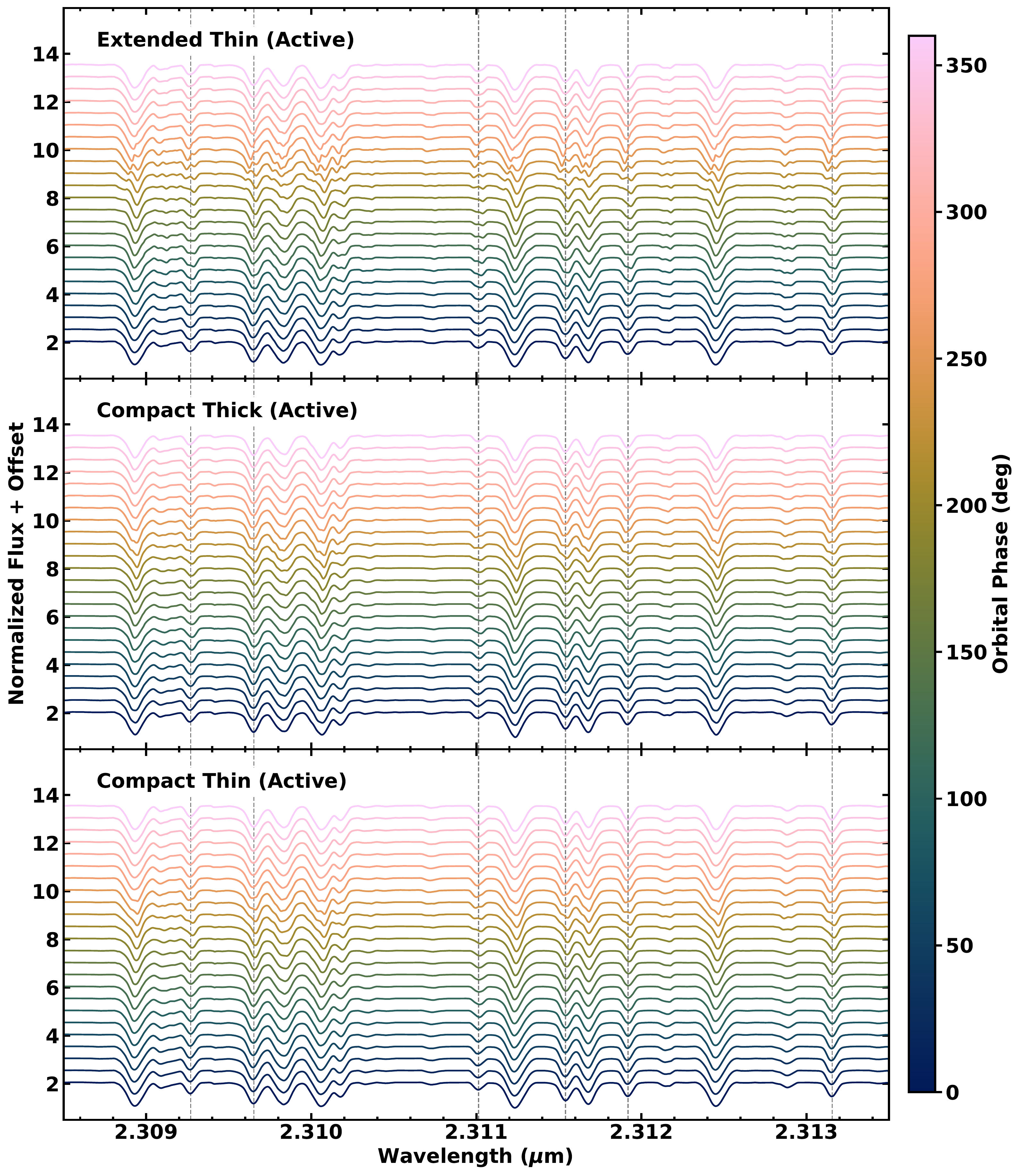}
    \caption{Same as in Figure \ref{fig:all_spectra1}, but for the extended thin cloud model (top), compact thick cloud model (middle), and compact thin cloud model (bottom), all with radiatively active clouds.}
    \label{fig:all_spectra2}
\end{figure*}

While the discussion of the main text focused on the results of the cloud-free and extended thick cloud models, here we present additional figures from the compact thin, compact thick, and extended thin cloud simulations (with active clouds). Temperature, wind, and cloud optical depth maps from each active cloud model are shown in Figure \ref{fig:all_temp_maps}, and corresponding temperature-pressure profiles and thermal inversion maps are shown in Figure \ref{fig:all_tp+inversion}. Finally, complete sets of the simulated Doppler-shifted spectra from each model are shown in Figures \ref{fig:all_spectra1} and \ref{fig:all_spectra2}. As summarized in Section \ref{sec:methods:gcm}---and fully described by \citet{roman+2019}---``\emph{thick}'' refers to GCM runs which assumed 1/10 of the gaseous cloud species condensed, while ``\emph{thin}'' refers to models in which 1/100 of the gas condensed. Likewise, ``\emph{extended}'' refers to GCM runs in which clouds extended from the cloud base pressure to the 0.1 mbar pressure level, and ``\emph{compact}'' refers to models in which the vertical extent of clouds was limited to $\sim$1.4 scale heights above the cloud base pressure.

\end{document}